\documentclass[journal,10pt,letterpaper]{IEEEtran}
\pdfoutput=1

\usepackage[hyphens]{url}
\usepackage{float}
\usepackage{placeins}
\usepackage{stfloats}

\usepackage[utf8]{inputenc}
\usepackage{color}
\usepackage{pifont}
\newcommand{\circleONE}{\text{\ding{172}}}
\newcommand{\circleTWO}{\text{\ding{173}}}
\newcommand{\circleTHREE}{\text{\ding{174}}}

\usepackage{diagbox}
\usepackage{minibox}	 
\usepackage{hhline}
 
\usepackage{enumitem}

\usepackage{verbatim}
\usepackage{cite}
\usepackage{graphicx}
\usepackage[caption=false]{subfig}
\usepackage{epsfig}
\usepackage[cmex10]{amsmath}
\usepackage{bm}
\usepackage{amsthm}
\usepackage{thmtools,thm-restate}
\usepackage{amssymb}
\usepackage{mathrsfs}
\usepackage{cases}

\usepackage{algorithm}
\usepackage[spaceRequire=true,indLines=false]{algpseudocodex}

\usepackage{array}
\usepackage{multirow}
\usepackage{makecell}

\newcommand{\argmin}{\operatornamewithlimits{arg\ min}}
\newcommand{\argmax}{\operatornamewithlimits{arg\ max}}

\usepackage[bottom]{footmisc}
\usepackage{tablefootnote}

\usepackage{pifont}


\usepackage{cleveref}
\crefformat{section}{\S#2#1#3}
\Crefformat{section}{\S#2#1#3}

\AtBeginDocument{
	\abovedisplayskip=4pt plus 3pt minus 3pt
	\abovedisplayshortskip=4pt plus 2pt minus 2pt
	\belowdisplayskip=4pt plus 3pt minus 3pt
	\belowdisplayshortskip=4pt plus 2pt minus 2pt
}

\IEEEoverridecommandlockouts

\begin{document}

\title{Socially Aware V2X Localized QoS}

\author{	
	\IEEEauthorblockN{Rafael Kaliski ~\IEEEmembership{Senior Member, ~IEEE}}
	\IEEEauthorblockN{Yue-Hua Han}
	
	\thanks{This work's funding support is from  
    Taiwan's National Science and Technology Council (NSTC), grants:
    NSTC 108-2218-E-011-036-MY3 an NSTC 112-2221-E-110-023-}
	\thanks{Rafael Kaliski (corresponding author) is with the Department of Computer Science and Engineering, National Sun Yat-sen University, Republic of China, Email: rkaliski@ieee.org}
	\thanks{Han Yue-hua is with the Research Center for Information Technology Innovation, Academia Sinica and also with the
	Department of Computer Science and Information Engineering, National Taiwan University of Science and Technology, Republic of China,
	Email: vhanxtreme@gmail.com} 
    \thanks{This work has been submitted to IEEE for possible publication. Copyright may be transferred without notice, after which this version may no longer be accessible.} 
}

\maketitle 

\begin{abstract}
 Vehicle-to-everything (V2X) is a core 5G technology. V2X and its enabler, Device-to-Device (D2D),  are essential for the Internet of Things (IoT) and the Internet of Vehicles (IoV). V2X enables vehicles to communicate with other vehicles (V2V), networks (V2N), and infrastructure (V2I). While V2X enables ubiquitous vehicular connectivity, the impact of bursty data on the network's overall Quality of Service (QoS), such as when a vehicle accident occurs, is often ignored. In this work, we study both 4G and 5G V2X utilizing Evolved Universal Terrestrial Radio Access New Radio (E-UTRA-NR) and propose the use of socially aware 5G NR Dual Connectivity (en-DC\footnote{Per Non-Standalone 5G\cite{TS_37340}, in E-UTRA-NR Dual Connectivity (en-DC) 5G New Radio (NR) and 4G radios coexist and are connected to a 4G core. UEs may connect to 4G and 5G Base Stations (BSs) simultaneously or choose to only connect to 4G BS. En-DC enables legacy technology to utilize 5G-enabled NR deployments.}) for traffic differentiation. We also propose localized QoS, wherein high-priority QoS flows traverse 5G road side units (RSUs) and normal-priority QoS flows traverse 4G Base Station (BS).

 We formulate a max-min fair QoS-aware Non-Orthogonal Multiple Access (NOMA) resource allocation scheme, QoS reclassify. QoS reclassify enables localized QoS and traffic steering to mitigate bursty network traffic's impact on the network's overall QoS. We then solve QoS reclassify via Integer Linear Programming (ILP) and derive its approximation. We demonstrate that both optimal and approximation QoS reclassify resource allocation schemes in our socially aware QoS management methodology outperform socially unaware legacy 4G V2X algorithms (no localized QoS support, no traffic steering) and socially aware 5G V2X (no localized QoS support, yet utilizes traffic steering). Our proposed QoS reclassify scheme's QoS flow end-to-end latency requires only $\approx 15\%$ of the time legacy 4G V2X requires.
\end{abstract}		


\IEEEpeerreviewmaketitle

\section{Introduction}
\label{sec:Intro}
	Vehicle-to-everything (V2X) is a core 5G technology. V2X and its enabler, Device-to-Device (D2D), are essential for the Internet of Things (IoT) and the Internet of Vehicles (IoV). V2X enables vehicles to communicate with other vehicles (V2V), infrastructure (V2I), and networks (V2N). Vehicular network traffic consists of low-latency, high-priority, typically safety-related, and normal-latency normal-priority Quality of Service (QoS) flows\cite{5GAA, 3gpp2018service}. Due to the heterogeneous networks and the multitude of services demanded by V2X (such as live video, road conditions warning systems, internet, and traffic information), coupled with the massive number of connected vehicles, meeting system-wide QoS requirements remains a challenge~\cite{Vehicular_networks_smart_roads}. V2X services often have strict latency and reliability requirements. Due to the high number of vehicles on the road and critical safety-related network traffic, V2X must meet massive Machine Type Communication (mMTC) and Ultra-reliable low-latency communication (URLLC) requirements.
	
 	Combining Road Side Units (RSUs) and Non-Orthogonal Multiple Access (NOMA) enables higher vehicle density via spectrum reuse. RSUs provide localized information, while NOMA enables more efficient use of the available spectrum by exploiting spatial diversity\cite{MAX_SUM_AND_MAX_MIN_PF_NOMA} of the allocated vehicles. Ensuring the QoS for vehicular and non-vehicular traffic remains challenging due to localized hotspots of vehicular data exchange, such as at busy intersections, rush hour on a highway, and areas prone to traffic accidents\cite{5GAA}. High-priority safety-related V2X network traffic is inherently bursty and can have a payload ranging from 190 to 1100 bytes per the 5G Automotive Association \cite{5GAA} (5GAA) and $3^{rd}$ Generation Partnership Project\cite{TR_38886} (3GPP). When many vehicles transmit network traffic close in time and to the same macrocell Base Station (BS), the overall system-wide QoS of non-safety V2X traffic can inadvertently and unnecessarily be impacted. 
 	
    \begin{figure}[t]
    	 \centering 
    	 \frame{\includegraphics[trim=0mm 0mm 0mm 0mm,clip,width=0.8\columnwidth]{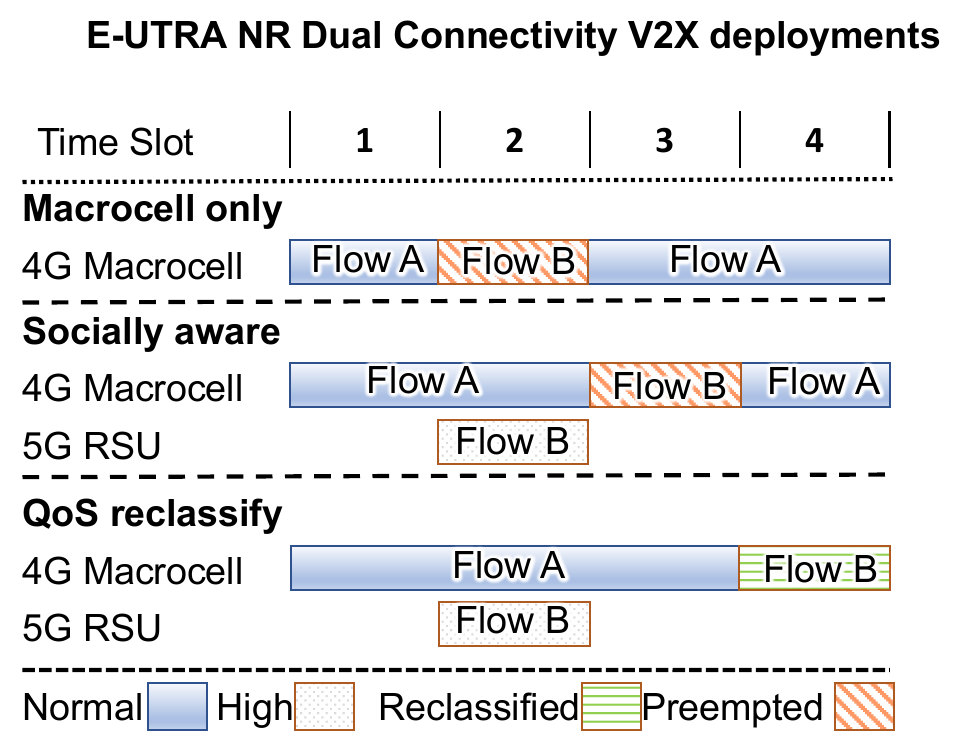}}
    	 \caption{Network traffic end-to-end QoS flow latency example with 4G LTE macrocell BS and 5G NR RSU Dual Connectivity (en-DC). RSUs provide localized resources, while macrocell BSs provide regional resources. \textit{Macrocell only} does not use traffic steering nor QoS reclassification. \textit{Socially aware} uses traffic steering only. \textit{QoS reclassify} uses traffic steering and QoS reclassification, where vehicular originating high-priority traffic is transmitted to the RSU and then forwarded to the macrocell BS for global dissemination. \textit{QoS reclassify} reclassifies high-priority traffic as a normal-priority before forwarding a QoS flow from an RSU to a macrocell BS.  In contrast, normal-priority traffic is directed only to the macrocell BS.}
    	 \label{fig:QoS_ReMap}
	\end{figure}
	
	When a vehicle generates high-priority / critical traffic, such as when a traffic accident occurs or a request for help is necessary, the network traffic is considered exigent to the vehicles in the vicinity. As the vehicle may not know the target audience or network destination, the exigent traffic is necessarily broadcast over the network using the Road Condition Warning System (RCWS) traffic type as high-priority network traffic. 
    If we only consider network conditions, addressing end-to-end network flow latency in 5G V2X networks can prove difficult due to the data's need to traverse the network core and meet both 5G New Radio (NR) and Long Term Evolution (LTE) QoS latency requirements over the heterogeneous network. To alleviate a network flow's impact on QoS in 5G V2X, we propose adding social awareness and QoS-Reclassification to 5G V2X networks/endpoints. Socially aware networks consider the origin and destination of the network traffic. In socially aware networks, the network flows are steered toward different macrocell BSs per network flow requirements. QoS-Reclassification consider locality and priority, high-priority QoS flows only remain within the locality of their origin before being forwarded and reclassified as normal-priority.
    
    For example, compare the three V2X deployments in Fig. \ref{fig:QoS_ReMap}.  Our example focuses on 4G LTE macrocell and 5G NR RSU Dual connectivity (en-DC) to enable more efficient downlink resource usage. Each vehicle connects to the closest in-range RSU and the closest BS. The RSUs provide localized resources, while macrocell BSs provide regional resources. We consider QoS Flow A a normal-priority QoS flow, three timeslots in duration, that always uses a BS. Next, we consider Flow B a high-priority, single-time slot in duration QoS flow, which may use either an RSU or a BS. The primary difference between the three deployments for Flow B is traffic steering and QoS reclassification. In macrocell-only deployments, QoS flow B uses the macrocell BS. In \textit{socially aware} deployments, QoS Flow B is directed to an RSU and forwarded to a macrocell BS. In contrast, in the \textit{QoS reclassify} deployment, QoS Flow B is not only directed to an RSU but is also reclassified from high-priority to normal-priority prior to being forwarded to the macrocell BS.   
    
	Based on the example above, QoS reclassification and network traffic differentiation can help mitigate the impact on QoS due to heterogeneous traffic types and congestion. Since Flow B is not exigent to the macrocell BS, it no longer needs to preempt Flow A on the macrocell BS after the BS reclassifies it as a normal-priority. This paper proposes socially aware traffic steering and QoS reclassification of V2X resource allocation mechanisms. We compare these mechanisms against a macrocell-only V2X resource allocation mechanism used in standard 4G V2X.
 
    Socially aware V2X implies that the network considers the originating, intermediate, and destination nodes when determining if a QoS flow's priority is necessary to achieve the QoS flow's localized requirements for each network segment. One method of extracting localized network segment-specific QoS flow requirements is to consider the traffic type and overlay the social network (path from the transmitter to destination group/users) on top of the network topology to create a socially aware network \cite{vegni2015survey} topology. Socially aware networks can account for network traffic relevance for end-users and adjust QoS accordingly for each network segment.
    This methodology enables us to dynamically adjust a QoS flow's settings to minimize a QoS flow's adverse impact on other QoS flows while still meeting the original QoS flow's requirements. As a transmitter may reside on one network, yet the receiver(s) on another, this problem may be formulated as a heterogeneous network QoS flow end-to-end timing closure problem. QoS Flow end-end-end time is calculated from the first packet generation to the last packet received.
    This research considers multiple non-overlapping RSUs and overlapping macrocell BSs utilizing QoS-aware traffic steering in the context of 5G non-standalone dual-connectivity networks.

	\subsection{Contributions}
    Prior research did not consider how social awareness and the network traffic type are interrelated and play an integral role in traffic-differentiated BS/RSU selection. Neither did previous works consider the impact of high-priority localized QoS flows on normal-priority regional QoS flows, i.e., QoS priority was not considered a localizable property.
	In this paper, we expand upon our previous work\cite{V2X_NOMA_DC}, provide insight into, and analyze socially aware V2X QoS for NOMA-enabled\footnote{Our proposed scheme can work with Orthogonal Frequency Division Multiple-Access (OFDMA), yet without NOMA only a lower vehicular density is possible.} V2X with Dual-Connectivity (DC). In addition, we provide a more detailed network architecture and a more representative of cellular networks' max-min fair QoS-aware water filling NOMA-enabled resource allocation algorithm and its approximation algorithm.  
	
	The main contributions of this work are:

	\begin{itemize}
        \item We expand on our previous work\cite{V2X_NOMA_DC} and discuss socially aware dynamic QoS determination and its impact on end-to-end\footnote{In this paper, we consider QoS Flow end-to-end latency as it better reflects the impact due to QoS flow preemption. We only consider radio transmission and queuing time from the transmission of the first packet of a QoS flow until the reception of the last packet. In our analysis, we consider the back-end transmission time negligible.} latency in heterogeneous networks. We also discuss the network architecture required to support the proposed QoS reclassification NOMA scheme via 4G macrocell BSs, 5G RSUs, and co-located Multi-access Edge Computing (MEC) servers.
        \item We consider network broadcast of QoS flows (safety-related) and the more complex scenario of groupcast communications (non-safety related) between different BSs. In addition, we extend traffic analysis to include multiple traffic distributions. Then, we investigate the time-out ratio and queuing time from both a broadcast and multicast perspective.
        \item Based on the queuing analysis, we propose a max-min fair QoS-aware water-filling mechanism as necessitated by safety/critical network flows. Our prior work only used a max-min QoS-aware water-filling mechanism. As such, edge devices were more prone to starvation. 
        \item We derive a tractable approximation algorithm utilizing a maximum benefit and minimum harm heuristic based on the proposed water-filling mechanism. Our mechanism selects candidate Resource Block (RB) locations in a manner that provides the most benefit to the current group while minimizing the interference to other groups / future RB allocations. Using Simulation of Urban Mobility (SUMO), we compared the proposed approximation algorithm to the optimal algorithm in a real-world scenario.
	\end{itemize}
	
	In \cref{sec:RW}, we discuss related works. Then, in \cref{sec:SysModel_ProbStatement}, we present the system model and the problem formulation. After which, in \cref{sec:Ana_V2X_SN}, we analyze the network queues, NOMA, and their relationship to 5G QoS Flows. Afterward, in \cref{sec:MMFQF}, we present the max-min fair QoS-aware water filling mechanism and its approximation in \cref{sec:Approx_Alg}. Finally, in \cref{sec:Sim_Dis}, we present our simulations and discuss the results. Then we conclude in \cref{Sec:Conclusion} and discuss future work in \cref{sec:Future_Work}.

\section{Related Works}
\label{sec:RW}

	Common LTE V2X resource allocation methods \cite{masmoudi2019survey} attempt to address resource allocation by finding a balance between V2V, V2I, and non-vehicular traffic. Prior research utilized hypergraph matching \cite{V2X_HYPERGRAPH_NO_NOMA} to determine how to share resources among vehicles and cellular users. 
	Accounting for resource utilization and interference alone limits the achievable solutions. Thus, in recent years, resource allocation has considered social awareness.
    As D2D is an essential foundation of socially aware communication for next-generation networks, the authors 
	in \cite{enLTED2Dproxservice} investigated the system-level performance of D2D proximity services (ProSe) for two-hop User Equipment (UE) relays with priority control proposed in 3GPP Release 13.
 
    Then, in \cite{SocialAwareV2XUnderlayingCellular}, the authors focus on V2I/V2V resource allocation with social awareness.
    The goal was to maximize the sum capacity of V2I links while also providing reliable V2V links via the proposed socially aware clustering resource allocation (SACRA) algorithm. SACRA consisted of a socially aware V2V link clustering algorithm that minimizes the intra-cluster interference of each social community. A matching theory resource allocation algorithm was derived. The works above do not consider dynamic QoS, where the QoS flow's settings are local to a subset of the network flow's path.
    
	One of the main pillars for V2X communication and 5G is URLLC. Unfortunately, an aspect of URLLC is that it is particularly susceptible to interference. \cite{IARR_5G_URLLC} formulated the inter-cell Interference-Aware Radio Resource (IARR) allocation problem to address interference, and derived a progressive heuristic solution to improve link reliability in multi-cell cellular networks. Another approach, proactive cell association \cite{ULL_MN}, uses a virtual cell associated with multiple access points simultaneously. Using open-loop communication increased the reliability of the device's connection while decreasing latency. In \cite{NOMAtoNR}, the authors improve the system sum capacity of the NOMA-integrated NR V2X system via a two-phase approach. The first phase provides a centralized graph-based sub-channel allocation for each vehicle group. Then, in the second phase, a distributed cooperative game approach controls NOMA-based vehicle group communication power.  
    Finally, to address the needs of URLLC in V2X, \cite{C_NOMA} proposed using relay-assisted NOMA to mitigate traffic congestion and reduce latency via power control.
    These URLLC works address interference and congestion yet do not consider the network side of QoS, i.e., how the network allocates resources among QoS flows or QoS flow prioritization. 
	
	Modern V2X systems use RSUs, enabling low-latency communications / localized information for vehicles. However, LTE-based V2X is unlikely to address the high number of high-priority connections necessitated by V2X and its associated bursty URLLC-related traffic safety requirements in a QoS impact-minimizing manner. The network must meet the diverse QoS requirements of multiple vehicular users\cite{5GAA, Vehicular_networks_smart_roads}. Considering the high volume of bursty network traffic typical to scenes of accidents, the unintentional adverse impact on normal-priority QoS flows becomes evident. The high aggregate volume of high-priority messages causes the end-to-end time of normal-priority traffic to increase. 
    In recent years, attempts to address the high volume of connections include using NOMA\cite{Tutorial_NOMA_5G} in V2X network designs. NOMA can increase radio spectral efficiency by exploiting spatial diversity, i.e., the distance between each vehicle and the BS. NOMA also enables Multi-User Diversity (MUD) by permitting the same radio RB to be utilized by multiple users who exhibit sufficiently different radio conditions/distances relative to the BS. In \cite{COOP_NOMA_BCMC_LLHR_V2X}, vehicles can only communicate indirectly with the macrocell BS via RSUs, i.e., each RSU acts as a layer-3 relay. The benefit, as determined by this research work, was that using full-duplex NOMA-enabled relays enabled better QoS and higher user fairness than only half-duplex relays. NOMA resource allocation efficiency increases when the BS's resource allocation is locally adjustable. For more efficient V2X grouping and higher spectral efficiency \cite{INTF_HYPERGRAPH_NOMA_V2X} uses D2D and NOMA. D2D enables a group to reuse the same resources as other groups in a non-orthogonal localized manner, while NOMA enables more efficient intra-group resource allocation. These prior works increased the number of simultaneously supported users yet did not consider how localized bursty traffic impacts QoS flow latency.

	Regarding traffic differentiated routing, \cite{Traffic_Diff_Clustering_Routing} analyzed delay-sensitive services and their QoS requirements in a V2X multi-hop network. Based on the tradeoff between directing traffic via Dedicated Short-Range Communications (DSRC) to either RSUs (low-bandwidth, long end-to-end delay) vs. using cellular connections, cellular networks incur bandwidth costs yet have a higher bandwidth and a lower end-to-end delay. Per this insight, the authors proposed a traffic-differentiated clustering routing mechanism in a Software Defined Network (SDN)-enabled hybrid vehicular network. The results showed that for a set bandwidth cost, when the cellular latency is higher than the multi-hop DSRC latency, vehicles near RSUs will choose DSRC.
	When considering localized routing, \cite{Traffic_Diff_Localized_Routing} analyzed and proposed a Wireless Sensor Network (WSN) localized QoS routing protocol. In the protocol, each packet attempts to meet the required data-related QoS metrics while also considering power efficiency. The proposed protocol ensures a tradeoff between energy consumed and the achieved QoS metrics. In other words, the protocol achieves low battery consumption for low to average critical packet rates. At the same time, at higher loads, high reliability is still maintained.  
    When considering suitable interfaces for communication, \cite{FUZZRA} used fuzzy logic to determine the suitability of a link's side link resources. The resources were either reserved or managed, as determined by the BS, to maximize resource reuse and improve network performance.
    Nevertheless, these works do not consider dynamic QoS.  Nor did they consider QoS-differentiated BS selection.

    When considering QoS-aware protocols, we consider systems designed to optimize the overall QoS. For example, \cite{Topo_Link_QA_Geographic} uses topological and link-quality aware geography to derive an opportunistic routing protocol to determine the selection of the next forwarding node to improve QoS. When considering IoT applications, \cite{IoT_App_Fog} proposed a generic IoT framework for deploying QoS-aware multi-component IoT applications over fog infrastructures.
    In terms of vehicle networks, \cite{QoS_Content_Cellular_V2V} considers the social relationship between V2V users, link reliability, and QoS of V2N and V2V users to maximize the weighted sum rate of V2V users.
    In Vehicular Ad-hoc NETworks (VANETs) \cite{VANET_routing} used ant colony optimization to compute feasible routes subject to multiple QoS constraints dependent on the network traffic type.
    The works mentioned above do not account for QoS localization wherein the network flow's exigency depends on the recipient and their relationship to the event's originator. These methods do not consider the impact of unnecessarily non-exigent high-priority events on normal-priority QoS flows.

    In recent years, Intelligent Reflective Surfaces (IRSs) and beamforming have become attractive means of improving signal quality in VANETs. IRSs enables the reflective surface to play a role in the communication system, enabling optimization for different QoS goals \cite{IRS_6G} such as higher rates, enlarged coverage area, and uninterrupted connectivity. For example, in \cite{pan2022artificial}, IRSs using deep learning and beamforming was proposed to improve signal quality while considering energy efficiency. While this method improves the radio performance of V2N communication, this method requires additional infrastructure to aid in signal boosting. 
    \cite{xie2021reflexive} optimized user grouping and BS blocklength allocation while performing IRS beamforming to reduce communication latency in a URLLC downlink system. 
    None of these works address the QoS impact that high-priority QoS flows have on normal-priority QoS flows.  

	Our work bridges the gap between socially aware networks and traffic-differentiated routing over regional and QoS-aware resources with localized QoS flows in 5G non-stand-alone V2X networks. Our QoS-aware methodology enables network flows to fairly compete for resources at a locality-level granularity per each QoS flow's exigency relative to each locality. Unlike the prior works, we consider how a network can intelligently, dynamically, and transparently adjust QoS settings of network traffic in a non-impacting manner while improving the latency experienced by lower-priority network flows.

\section{System Model and Problem Statement}
\label{sec:SysModel_ProbStatement}

QoS flow management is problematic for V2X communications\cite{Vehicular_networks_smart_roads} due to the differing technologies, architectures, network topologies, QoS definitions, network connectivity, channel quality, target group identification (determining the network flow's destination), and stringent end-to-end QoS guarantees. 
QoS requirements can be even more challenging in a high volume of vehicular-originating network flows. In V2X networks, high network flow volume commonly occurs at bottlenecks/accident points in the road network. 
Efficient QoS-aware resource management requires real-time group membership and location information to reduce QoS flow end-to-end latency and derive more effective V2X socially aware QoS-based resource allocation methodologies via \textit{socially aware networks}. 

\begin{figure}[t]
	\centering 
	\frame{\includegraphics[trim = 0mm 0mm 0mm 0mm, clip, width=1\columnwidth]{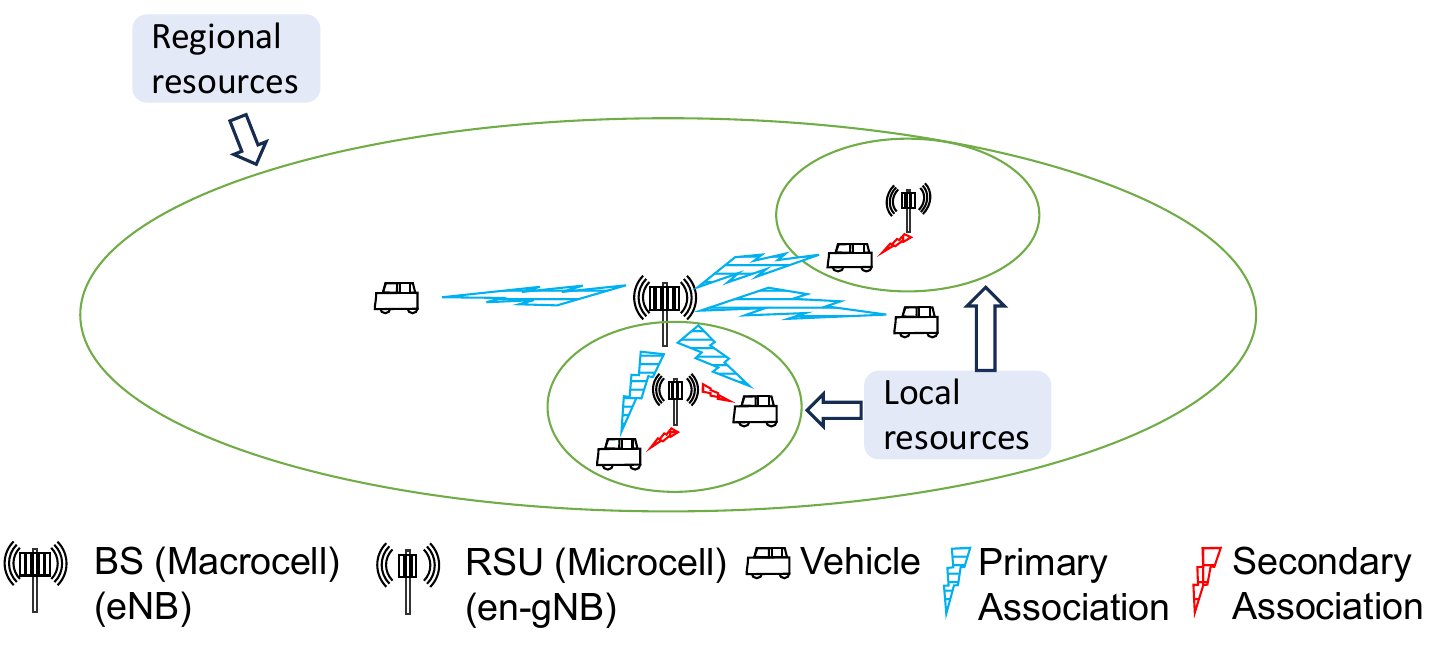}}
	\caption{V2X en-DC System Model: Every vehicle has a primary association with a BS and potentially a secondary association with an RSU.  The BS provides regional resources, while the RSU provides local resources.}	
	\label{fig:V2X_DC_System_Model}	
\end{figure}

Our V2X system model is shown in Fig. \ref{fig:V2X_DC_System_Model}. There are $|\textbf{R}|$ RSUs, each assigned $B_R$ Mhz bandwidth symmetrically divided between $B_R/2$ Mhz uplink and $B_R/2$ Mhz downlink, and with an effective range of $D_R$ (meters). Also, there are $|\textbf{B}|$ BSs, each assigned $B_M$ MHz bandwidth symmetrically divided between $B_M/2$ Mhz uplink and $B_M/2$ Mhz downlink, and with an effective range $D_M$ (meters).  
In addition, there are $|\textbf{U}_i|$ vehicles / UEs per the $i^{th}$ BS.
Each RSU is located at a road intersection and is within the coverage of a single BS (RSUs do not overlap.) The RSUs and BSs operate on orthogonal spectra, yet RSUs and BSs share the spectra with other RSUs and BSs. RSU and BS utilize NOMA due to high vehicular traffic and limited bandwidth. Each vehicle maintains a link to a BS at all times. When a vehicle is within range of an RSU, it also automatically associates with the RSU with the strongest Signal-to-Interference-plus-Noise Ratio (SINR).
Note:  In this research, sets are denoted in bold. 4G macrocell BSs are referred to as BSs.  5G RSUs are referred to as RSUs.

\begin{figure}[t]
	\centering
    \frame{\includegraphics[trim = 0mm 0mm 0mm 0mm, clip, width=.9\columnwidth]{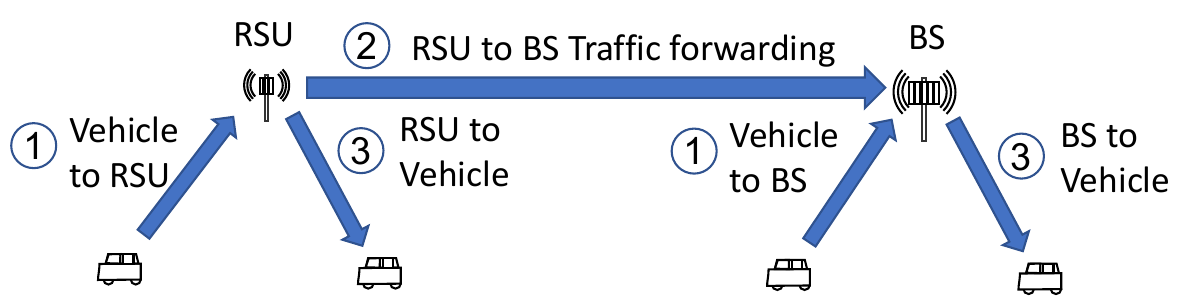}}
	\caption{V2X en-DC Network Segments: Network segment $\circleONE$ provides uplink resources, $\circleTWO$ provides inter-AP connectivity for QoS flow forwarding, and $\circleTHREE$ provides downlink resources.} 	
	\label{fig:V2X_Net_Segs}	
\end{figure}

For our system, per Fig. \ref{fig:V2X_Net_Segs}, there are three potential segments for a QoS flow between the originating transmitter and destination receiver.  
The first segment $s_i = \circleONE$ is between a transmitting vehicle and its associated RSU/BS. The second segment $s_i = \circleTWO$ is from an RSU to a BS. The third segment $s_i = \circleTHREE$ is between the transmitting RSU/BS and the target vehicle. Note: the second segment only exists in a network flow that traverses an RSU to BS link to reach its destination. As a result, for multi-destination network flows, each QoS flow comprises two or more network segments.

The goal of our V2X system is to minimize the overall end-to-end QoS flow latency over the aggregate set of QoS flows per the QoS group $\mathbf{G}_p$, the set of all QoS Priorities $\mathbf{P}_g$, subject to the assigned QoS priorities $p_{g, i}$. Our V2X system considers all intermediate network segments in  $\mathbf{S}_g$ traversed by the QoS flow $g$ between the origin and set of terminal recipients, as shown in \eqref{eq:S_obj}. In other words, in our proposed system, the QoS priority is a network segment variable we optimize over to minimize unnecessary preemption and localize each QoS flow's priority. %
\begin{align}
    \label{eq:S_obj}
    \argmin_{\mathbf{x}_g} &\sum_{g \in \mathbf{G}_p} \sum_{i \in \mathbf{S}_g} \Delta({s_i,p_{g,i}}) + \rho_{i} \text{ } \forall p_{g,i} \in \mathbf{P}_g \\
    &\mathbf{P}_g = \begin{cases}
    p_h \text{ } (x_{g,i} = 1) \\
    p_n \text{ } (x_{g,i} = 0) 
    \end{cases} \nonumber \\
    &\mathbf{S}_g = \{s_1, s_{2}, ... s_N\},  \mathbf{x}_g = \{x_0, x_1, ... x_N\} \nonumber 
\end{align}
where $\Delta({s_i},p_{g,i})$ represents the delay incurred by a QoS Flow $g$ per network segment $s_i$ per local queuing delay, given priority $p_{g,i}$, and $\rho_{i}$ is the processing delay incurred by the transmitter.  $p_{g, i}$ takes the value of $p_h$ if the transmitter-receiver grouping indicates that the traffic is exigent ($x_{g, i} = 1$), as high-priority QoS flow preempts normal-priority QoS flows. If the network traffic is not exigent ($x_{g, i} = 0$), the priority of the QoS flow is assigned $p_n$, i.e., normal-priority. Normal-priority QoS flows do not preempt other QoS flows.

\subsection{Social Awareness in Networks}
In an end-to-end QoS flow management design, we consider each QoS flow's priority and latency dependent on its associated social group (a social group refers to the QoS flow sender and its target recipients), its destination(s), and its QoS priority. In other words, a socially aware network considers QoS settings per network flow per network segment and each QoS flow's exigency to each recipient. Thus, each vehicle implicitly requests a social group by identifying its QoS flow's target recipients. The target recipients are, therefore, considered a user-controlled yet network-managed list of members. We note that the list of members may be implied (broadcast) or explicit per a user-defined group of recipients (unicast/groupcast). The network manages this list by tracking the target's and recipient's network locations. As such, the network is better equipped to address each QoS flow's requirements based on the exigency of the QoS flow to the network segment's locality. Note: we define a social group's QoS as being determined by the traffic's type, origin, destination, and initial QoS settings. Thus, a socially aware network tracks each QoS flow's destination and priority.

\subsection{System Architecture}
\begin{figure}[t]
	\centering
	\frame{\includegraphics[trim = 0mm 0mm 0mm 0mm, clip, width=0.9\columnwidth]{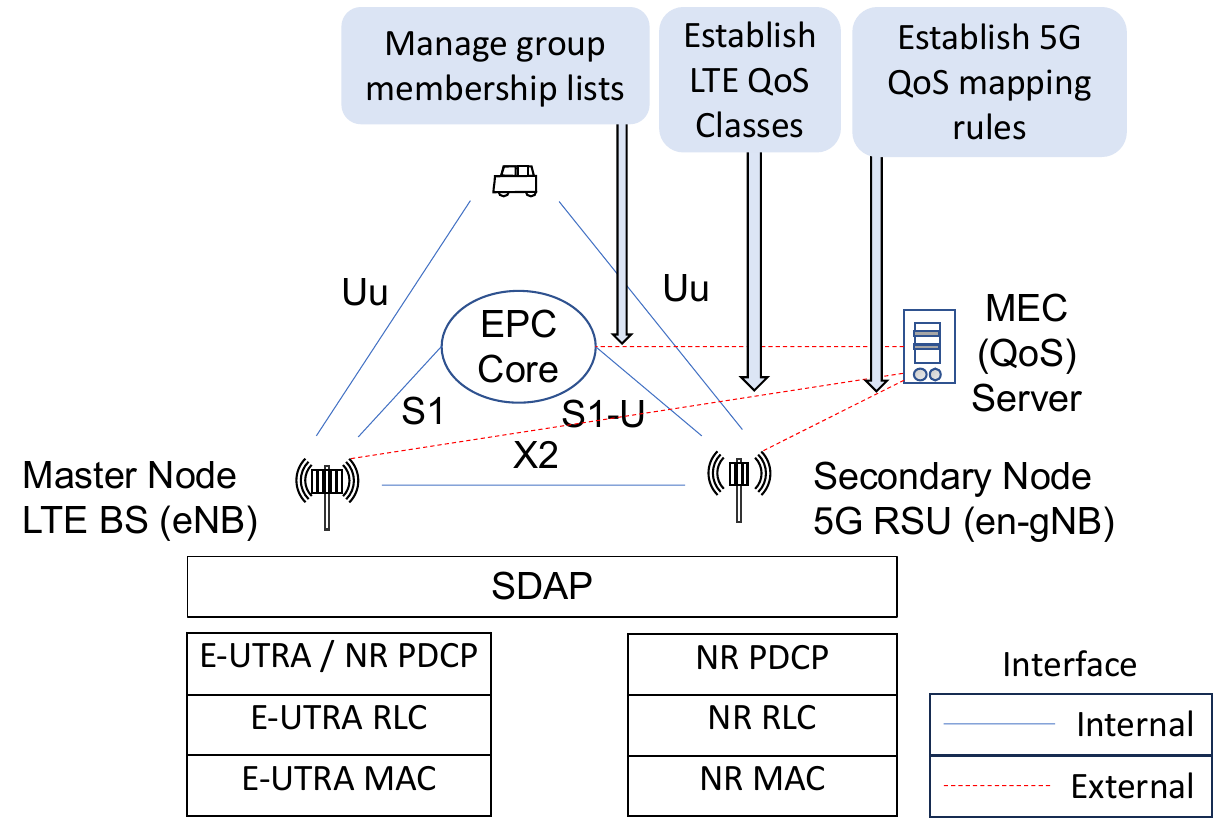}}
	\caption{V2X en-DC System Architecture: V2X en-DC uses an LTE BS, which acts as the master node, and a 5G NR RSU, which acts as a secondary node.  Both nodes terminate on the EPC core.  The MEC QoS server establishes QoS classes for the LTE BS, 5G QoS mapping rules for the 5G NR RSU, and performs group membership list management via the EPC core connectivity.} 		
	\label{fig:En_DC_System_Model}
\end{figure}

When considering social awareness and cellular traffic's QoS management, the relevance and exigency of the data effectively determine the QoS utilized by each BS/RSU. For example, safety-related messages are more relevant to the vehicles near the incident. As such, an event-originating vehicle would preferentially select an RSU over a BS as RSU deployments are local, i.e., less competition for wireless network resources, while BS deployments are regional. 
To achieve dual connectivity and to ensure the system remains compatible with 4G V2X, we utilize an en-DC\cite{TS_37340} system topology, as shown in Fig. \ref{fig:En_DC_System_Model}. 
In an en-DC system, the LTE BS  (eNodeB/eNB) is the Master Node (MN), while a 5G NR (enhanced gNodeB/en-gNB) RSU is considered the Secondary Node (SN).  
In en-DC, the BS and the RSU connect to the LTE Evolved Packet Core (EPC). 
For our topology, we associate each RSU with the closest BS  via an \textit{X2} interface. At the same time, each regional BS  (local RSU) connects to the EPC core via the \textit{S1} (\textit{S1-U}) interface. 

The Uu air interface connects each vehicle and its associated BS and RSU. For this DC setup, the upper layer's Service Data Adaptation Protocol (SDAP) function\cite{TR_37324} utilizes EPC-connected $3^{rd}$ party MEC servers which collect and maintain social group membership lists. The MEC servers are co-located with the EPC\cite{MEC_4G_5G} and modify the QoS priority and mapping rules. Vehicular originating uplink high-priority traffic is routed from the BS/RSU to the MEC. The MEC then broadcasts the high-priority traffic to the local BS/RSU, which then conditionally forwards the traffic. The RSU reclassifies the QoS flow as a normal-priority to the regional BS. For the 5G RAN (en-gNB), QoS mapping rules\cite{3gpp2020_5GS_arch} determine how network traffic, termed QoS flows in 5G, is mapped to data radio bearers. Via a QoS mapping rule, the RCWS traffic type is considered a high-priority QoS flow in both the uplink and downlink directions.   RCWS traffic is labeled high-priority via a QoS Class Identifier for LTE. The network traffic flow is set to normal-priority when forwarded between an RSU (5G) and the macrocell (4G).
The lower protocol layers, Packet Data Convergence Protocol (PDCP), Radio Link Control (RLC), and Medium Access Control (MAC), handle packet transmission and reception and are unaffected by the proposed localized QoS modification. This system design enables a vehicle to roam between different service providers as long as the service provider's network connects to the same or synchronized $3^{rd}$ party QoS server. The vehicle need not be concerned with how the QoS flow configuration changes as the SDAP can dynamically set the QoS properties for each 5G QoS flow per the network and the target social group. In other words, the user-specified QoS settings implicitly become a network-assisted setting. 

\section{V2X Queuing Times and Social Networks}
\label{sec:Ana_V2X_SN}
We now consider the impact of network traffic on QoS flow latency. 
Our proposed system comprises an $M/G/k$ queue (Markov arrivals driven by Poisson process, general distribution, k servers) associated with each radio interface.  As the backend, inter-AP, does not impose a limitation; the backend can be modeled as an $M/G/\infty$ queue, as there is close to zero delay ~\cite{li2019statistical}.

Regarding the radio interface, the number of concurrent QoS flows (each serviced QoS flow corresponds to a server) depends on the geometries associated with the NOMA RBs.  For 5G, high-priority QoS flows have a RB geometry of $360kHz \times 0.5ms$, while 5G normal-priority, or any 4G QoS flow, has a $180 kHz \times 1.0ms$ RB geometry.  For a bandwidth of $20 Mhz$ (4G BS), at $10\%$ overhead, there is $20Mhz \times 0.9 / 180kHz = 100 \times 2$ (bi-layer NOMA) $= 200 RBs$.  If each QoS flow only requires one RB, then the maximum $k = 200$ QoS flows for the BS.  For a bandwidth of $10 Mhz$ (5G RSU), with only high-priority RBs, at $10\%$ overhead, there is $10Mhz \times 0.9 / 360 kHz \times 2$ (bi-layer NOMA) $= 50 RBs$.  If each QoS flow only requires $1 RB$, the maximum $k = 50$ QoS flows for the RSU.

\begin{figure}[ht]
{\begin{minipage}{0.39\columnwidth}
\begin{align*}
    &Q_{j,s} = \frac{\lambda_{j,s} = \sum\limits_{t = 1}^m \mu_{i,t}}{\mu_{j,s}}  \\
    &Q_k = \frac{\lambda_k = \sum\limits_{s = 1}^n \mu_{j,s}}{\mu_k = \sum\limits^p_{r=1} \mu_{k,r}} \nonumber 
\end{align*}
\end{minipage}
\begin{minipage}{0.6\columnwidth}
    \centering
	\frame{\includegraphics[trim = 0mm 0mm 0mm 0mm, clip, width=\columnwidth]{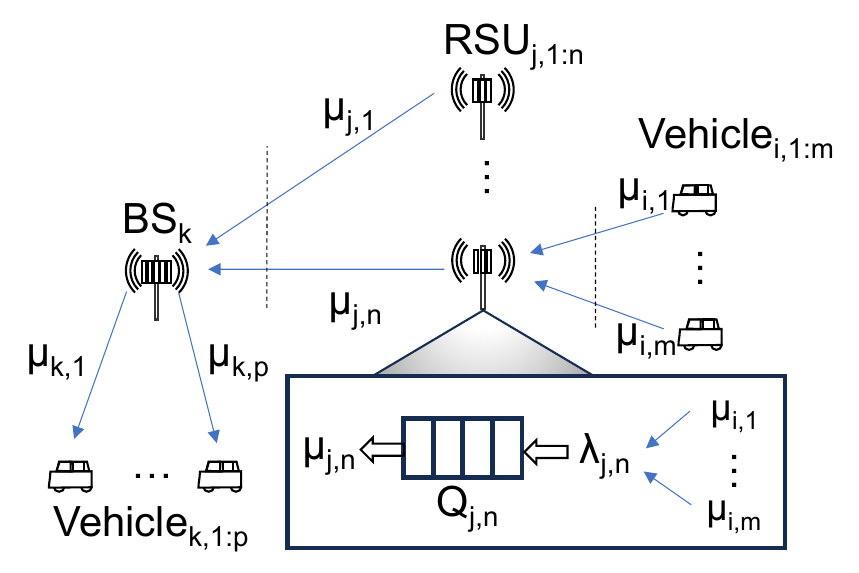}}	  
\end{minipage}}
\caption{UE-BS Queue model:  There are $m$ vehicles transmitting to RSU $j$, and $n$ RSUs transmitting to the BS $k$ which serves $p$ vehicles.  The individual data reception and transmission rates combine and split per a Poisson process.}
\label{fig:UE_BS_Queue}  
\end{figure}

For any BS/RSU $j$, when the network traffic transmission rate $\mu_j$ is greater than the reception rate $\lambda_j$, BS/RSU congestion does not occur as the network queue is considered stable. However, when the incoming network traffic exceeds the transmission rate, i.e., $\lambda_j > \mu_j$, the queue length and QoS flow latency increase due to BS/RSU congestion. In our research, the network queue length can increase due to too many UEs attempting to access the network simultaneously, a common occurrence in places with high vehicular traffic. It can also occur due to inter-BS network traffic data aggregation, such as when regional awareness of an event-generating RCWS traffic is required. Vehicles transmit high-priority events to the nearest RSU. We illustrate this scenario in Fig. \ref{fig:UE_BS_Queue}.  The regional BS is at level $k$, aggregating RSU-originating traffic at level $j$. Each aggregating vehicular traffic originates from level $i$.  $\mu_{i,t}$ represents a vehicle's traffic transmission rate and $\mu_{j,s}$ represents the outgoing traffic from the $s^{th}$ RSU.  $\lambda_k$ represents the incoming traffic rate from the set of RSUs $j,s$ to BS $k$ and $\mu_k$ represents BS $k$'s outgoing traffic rate.

When $Q_{j,s} > 1$, the vehicle-originating QoS flows incur additional latency via increased RSU transmit queuing duration. When $Q_k > 1$, the inter-BS traffic incurs additional latency via increased BS transmit queuing duration. In both cases, an increased QoS flow reception delay and higher QoS flow end-to-end latency occur. If the reception delay is too high, the network traffic is timed out and dropped.

In non-socially aware V2X networks, $Q_k = Q_j$ due to all data utilizing the macrocell BS. As such, $Q_j > 1$ can occur when high or normal-priority traffic exceeds the service rate, i.e., the BS transmission rate. In a socially aware dual connectivity network architecture, on the other hand, RSUs offer localized QoS, and BSs offer regional QoS. In other words, $Q_j$ is less likely to be greater than one as the uplink radio resources are only shared by the high-priority QoS flows. 
While socially aware networks can help ensure high-priority network traffic's QoS latency for local recipients, they inherently increase the network traffic bound for the nearest regional BS. Instead of queue congestion due to too many UEs attempting to access the network,  the increased traffic can cause queue congestion due to excessively high inter-BS network traffic, i.e., $Q_j < 1$, yet $Q_k > 1$. 
Consequently, the state of each queue impacts all BS traffic. Thus, the need for dynamic QoS reclassification becomes evident.  

When high-priority inter-BS traffic arrives at a destination BS, preempting normal-priority traffic, normal-priority traffic incurs a higher queuing time. By reclassifying the RSU originating high-priority traffic as normal-priority, the inter-BS traffic queuing time incurred and the corresponding increase in QoS flow end-to-end time latency can be mitigated.

\subsection{NOMA and 5G QoS Flows}

\label{sec:NOMA_5G_QoS}

For downlink power-domain NOMA deployments, the decoding order and interference experienced depend on the vehicle locations and allocation order of the same RB resources. Without loss of generality, for $N$ QoS flows, we assume a linear ordering of transmit powers $P_{i+1} \ge P_{i}$ (vehicle $i$ is the inner vehicle, vehicle $i+1$ is the outer vehicle.)
For a NOMA system where the outer vehicle's QoS flow is decoded prior to the inner vehicle's, Successive Interference Cancellation (SIC) can remove all higher-powered interfering signals. A vehicle allocated a QoS Flow with transmit power $P_i$ can be decoded by successively decoding and subtracting all signals with a transmit power higher than $P_i$. All signals with a transmit power lower than $P_i$ are deemed as interference and are not decodable. Therefore, NOMA-enabled vehicles can receive the composite downlink signal and the set of the corresponding transmit powers $\textbf{P}$ to determine which sets of overlapping transmissions should be decoded and discarded prior to obtaining the intended signal $S_i$, where $S_i = P_i |h_{ik}|^2$ ($h_{ik}$ corresponds to the transmission matrix for user $i$ in group $k$). We note that power-domain NOMA's improvement in spectral efficiency becomes more evident when the spatial diversity of the vehicles sharing common RB resources increases.

In QoS-based NOMA\cite{NOMA_IMPORTANCE_UNVELING_PF_QOS}, the decoding order ensures that the highest priority social group requiring a set of resources acquires the required bitrate over the assigned RBs. The BS allocates unused capacity to the next highest priority social group. In other words, for the highest priority social group with a set demand, the allocated QoS Flow bitrate $R_i$ must meet the below SINR requirement \eqref{eq:QoS_SINR_Req}.
\begin{align}
	SINR_i(P_i) = \frac{P_i |h_{ii}|^2}{\sum_{k=i+1}^N P_k|h_{ik}|^2+ \sigma_n^2} \label{eq:QoS_SINR_Req} \\
	B \cdot log_2 \left(1+SINR_i\right) \ge R_i,  
	c_i = f(SINR_i)  \nonumber \\
	\sum_{k=1}^N P_k \le 1, \text{ and } P_j \ge P_k, \forall j < k
	\label{eq:NOMA_Power_Frac}
\end{align}
where $N$ corresponds to the number of allocated QoS flows utilizing the same wireless channel / RBs. $B$ corresponds to the bandwidth allocated to the BS / RSU. $P_i$ corresponds to the transmit signal strength, and $h_{ik}$ corresponds to the weakest channel gain of all members in the social group $i$ who subscribe to the QoS flow $k$. The corresponding interference is determined by $P_k$ and $\sigma_n$. $P_k$ is the normalized power allocated to other QoS flows, and $\sigma_n$ is the Additive White Gaussian Noise (AWGN). The variable $c_i$ reflects the RB capacity as determined per the SINR $SINR_i$ to RB capacity mapping function $f(\cdot)$. The NOMA power coefficient constraints for the set of $N$ QoS flows using the same RB are per \eqref{eq:NOMA_Power_Frac}. For this research, we limit the number of power-domain multiplexed data streams per RB to 2, i.e., $|\textbf{P}| \le 2$. 

\subsection{QoS flow routing} 
In our proposed system, the social group of each vehicle's QoS flow depends on its network traffic type, origin, and destination(s). Assuming a vehicle is associated with a BS and an RSU, it directs its QoS flow toward either an RSU or a BS per the traffic type. In other words, 
the proposed system supports QoS flow differentiated services. Each social group effectively indicates its priority for associated QoS flows and each QoS flow's relevance and exigency. 

\begin{figure}
 \centering
 \frame{\includegraphics[trim=0mm 0mm 1mm 0mm, clip,width=1.0\columnwidth]{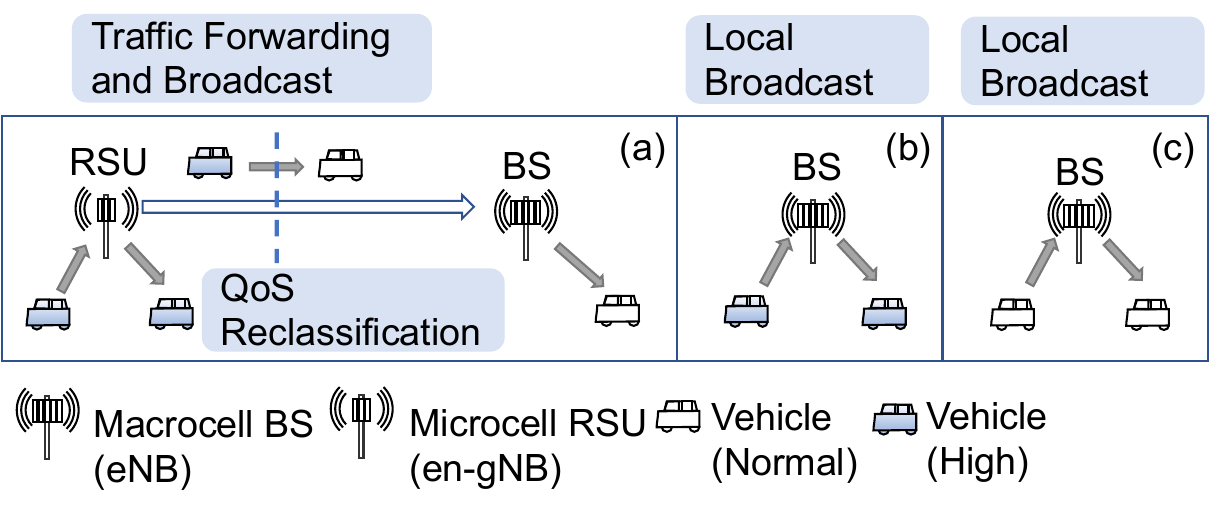}}
 \caption{QoS reclassification and traffic steering example: (a) RSU (en-gNB) originating high-priority QoS network flows are conditionally broadcast to the vehicles around the RSU. The network traffic is also QoS reclassified to normal-priority and then forwarded to the macrocell 4G BS eNB. (b, c) high-priority and normal-priority eNB-originating QoS flows are not QoS reclassified before being conditionally broadcast. Note: High-priority QoS flows out of range of any RSU are transmitted to the BS (b).}
 \label{fig:QoS_ReClassify}
 \vspace{-1.0\baselineskip}
\end{figure}

As shown in Fig. \ref{fig:QoS_ReClassify}, vehicles with critical / high-priority QoS flows transmit their data to the nearest RSU $(a)$. If no RSUs are within range, the vehicle transmits the QoS flow to the closest BS $(b)$. 
If the QoS flow is deemed relevant to the region, then it is forwarded by the RSU to the closest associated BS $(a)$. When the QoS flow is forwarded, the BS dynamically changes the QoS flow's social group to a normal-priority social group.
If the type of a QoS flow is normal-priority, it is transmitted directly to the BS $(c)$.    

This research investigates the broadcast RCWS and multiple groupcast scenarios, such as caravans/vehicle platoons with group video chat or other normal-priority shared data.

\subsection{Dynamic QoS}

Dynamic QoS aims to minimize the unnecessary delay incurred due to preemption by all QoS flows, i.e., minimizing $\Delta(s_i, p_{g,i})$ per \eqref{eq:S_obj}.  Consider a normal-priority QoS flow A of length 2, starting at time 0, and a high-priority QoS flow B of length 1, starting at time 1. Let the network segment’s delay based on the QoS flow A be $\Delta_A$, while the optimal delay achieved with QoS reclassification is $\Delta_A^*$.  Analyzing the three different scenarios:
\begin{itemize}
    \item \textbf{Receiver considers data exigent: } The QoS flow B remains high-priority when transmitting to the receiver. Thus, QoS flow B preempts QoS flow A, causing QoS flow A to incur delay ($\Delta_A=1$).  However, this delay is necessary ($\Delta_A^*=1$), i.e., $\Delta_A=\Delta_A^*$.
	\item \textbf{Receiver does not consider data exigent, no QoS reclassification: } The QoS flow B preempts QoS flow A, causing QoS flow A to incur delay ($\Delta_A=1$). The preemption was unnecessary as QoS flow B is not considered exigent by the receiver ($\Delta_A^*=0$), i.e. $\Delta_A>\Delta_A^*$.
	\item \textbf{Receiver does not consider data exigent, uses QoS reclassification: }  Since the receiver does not consider QoS flow B exigent, QoS flow B was reclassified as a normal-priority QoS flow, lowering QoS flow B’s impact to QoS flow A ($\Delta_A=0$).  QoS flow A did not need to be preempted by QoS flow B ($\Delta_A^*=0$), i.e., $\Delta_A=\Delta_A^*$.
\end{itemize}

In terms of dynamically adapting a QoS flow's priority, localizing QoS, we consider the impact of the number of groups and each group's size. A dilemma with constantly changing group size is the computational complexity of the resource allocation problem and how the network flows' QoS priority between the RSUs and BSs is adjusted to address the target group's QoS goal. The maximum number of groups, the number of QoS Flows, and the group size can vary vastly from one radio frame to another.  Thus a network aware dynamic QoS flow adaption methodology is necessary.

To determine the forwarding behavior and impact of a transmission queue due to a specific QoS flow, we need to know the population of the target group in a network segment.
If the target group is empty due to no member associating with the BS, then the RSU-originating flow is dropped by the BS, i.e., the BS's transmission queue does not grow. A prerequisite for target group size determination is that each BS must know each vehicle's location. Collecting location information is already performed as 4G and 5G support periodic and event-driven tracking area updates \cite{TS_36300} for paging purposes. A caveat with our QoS Flow forwarding system is that the backhaul's rate must exceed that of the radio transmission speed. Otherwise, the radio resources may become underutilized. For this work, we assume the backhaul's capacity and speed exceed those required by each BS and RSU.

\section{Max-Min Fair QoS-aware Water Filling}
\label{sec:MMFQF}
    Each BS's / RSU's RB geometry\footnote{For this research, as only high-priority / low-latency QoS flows use the RSUs, 5G NR RB geometry is limited to $360 kHz\cdot 0.5 ms$. LTE RB geometry is $180 kHz\cdot 1 ms$ per \cite{TR_38901}.} selection, resource assignment, and power coefficient allocation depend on each vehicle's Channel Quality Indicator (CQI\footnote{Detailed CQI reports\cite{TS_36213, TS_38214} include per RB / sub-channel SINR.}) report and the observed interference over every 180 kHz-wide segment of radio spectrum\footnote{For NR, the maximum capacity of each RB geometry is the same across all numerologies. Thus, selecting which geometry to use depends on the available resources and the latency requirements\cite{TR_38901}.}.
    Based on the CQI report and reported interference, the BS / RSU generates an SINR matrix to determine where each RB should be assigned. RB assignment determines RB geometry, frequency, and temporal location.  
    
    For our NOMA deployment, we consider bi-level NOMA.
    Regarding resource allocation and QoS, high-priority vehicles are first allocated QoS resources \cite{MAX_SUM_AND_MAX_MIN_PF_NOMA}. After that, the resource allocation mechanism allocates unused resources to normal-priority vehicles.
	The resource allocation problem considers multiple factors, including the group payload demand $g_\text{bits}$, the group size $g_\text{size}$, and the fairness $g_\text{fair}$ of a given group $g$. We formulate the max-min fair resource allocation problem, which maximizes the number of serviced users in \eqref{eq:MAX_MIN_FAIR}.
\begin{gather} 
    \textbf{P}^*: \argmax\min_g \left\lceil \frac{g_\text{bits}}{g_\text{size}g_\text{fair}} \right\rceil, \forall g \in \textbf{G}_p, \forall p \in \bm{\mathcal{P}} \label{eq:MAX_MIN_FAIR} \\
    g^{t}_\text{fair} = (1-\frac{1}{T_{c}})g^{t-1}_\text{fair} + \frac{g^{t}_\text{bits}}{T_{c}} \nonumber 
\end{gather}
    We define $\textbf{G}_p$ as the set of groups of priority $p$, $\bm{\mathcal{P}}$ as the set of priorities $\{p_h, p_n\}$, and the resource assignments as $\textbf{P}^*$.  Per max-min, we consider social groups with large group sizes or small demands as higher priority groups for a specific QoS priority. However, we also note that a pure max-min solution will result in demand accumulation, i.e., many high-demand groups will aggregate and eventually cause lower-demand groups to suffer from starvation. Therefore, we use max-min fair to scale the group size according to a control variable $g_\text{fair}$ that refers to the historical demands and changes at a convergence rate, $T_c$, as time elapses.  
    
    Unfortunately, this combinatorial optimization problem in \eqref{eq:MAX_MIN_FAIR} is non-convex. Thus, we derive a two-part optimal solution using an Integer Linear Programming (ILP) Block Coordinate Descent\cite{BCD} (BCD) approach that accounts for the impact of each NOMA RB allocation on future NOMA RB allocations. The original optimization problem can be decomposed into two sub-problems: (1.) the NOMA inner layer RB allocation sub-problem ($I_k$) and (2.) the NOMA outer layer RB allocation sub-problem ($O_k$). The resource allocation process first allocates resources to high-priority QoS flows, then normal-priority QoS flows. As the NOMA RBs are bi-layer, the 1st layer (inner) is allocated prior to the 2nd layer (outer).  The 1st layer may use up to the entire power budget, while the outer layer may only use the unused power budget.
    We optimize the resource allocation utilizing ILP for each allocation sub-problem, per \eqref{eq:TIME_VAR_FAIR_ALLOC}.
\begin{gather}
    \begin{gathered}
    \mathbf{x} = \sum_{c}^{C_{l,g}}{\sum_{i}^{L_g}{x_{g,l,c,i}}}\\
    x_{g,l,c,i} \in [0,1], l \in [0,1], \forall g \in \textbf{G}_p, p \in \bm{\mathcal{P}}, 
    \end{gathered}
    \label{eq:TIME_VAR_FAIR_ALLOC}
\end{gather}
    where $x_{g,l,c,i}$ indicates the assignment of a physical frequency-time RB $i$ to group $g$ in layer $l$ with Modulation and Coding Scheme (MCS\footnote{The corresponding MCS used by the BS is determined per each RB's CQI, as reported by each vehicle.}) $c$. 
    $C_{l,g}$ contains all possible MCS settings of the RB fraction. $L_g$ represents the number of different frequency-time RB locations within the radio resource allocation frame, e.g., for a RB of 180 kHz$\cdot$1 ms, there exist five different locations within a sub-frame of 900 kHz$\cdot$1 ms. The corresponding constraints are presented in \eqref{eq:RBF_PREVENT_OVERLAP} to \eqref{eq:RB_FIX_NOMA_LOW_PRIO}.

    Each 180 kHz$\cdot$0.5 ms RB fraction is constrained to only belong to a single RB in the same layer to prevent overlapping allocations\footnote{An RB's latency is determined by its geometry-specific numerology. Each numerology corresponds to a specific RB duration, RB bandwidth, and Sub-Carrier Spacing (SCS)\cite{TR_38901}. For this paper, normal-latency RBs use $15 kHz$ SCS, while low-latency RBs use $30 kHz$. 5G is capable of both low-latency (360 kHz$\cdot$0.5 ms) and normal-latency (180 kHz$\cdot$1 ms) RB geometries, while LTE is only capable of a normal-latency RB geometry.}. In addition, due to the nature of NOMA SIC, a single NOMA layer of an RB, may only be used by a single group at a time. Each RB must be assigned to different groups in each NOMA layer per \eqref{eq:RBF_PREVENT_OVERLAP}.
\begin{gather}
    \begin{gathered}
     \sum_{c}^{C_{l,g}}{\sum_{i}^{L_g}{  D(g,i,n)x_{g,l,c,i}}}  \le 1,\\
     l \in [0,1], n \in [1,N_{fr}], \forall g \in \textbf{G}_p, p \in \bm{\mathcal{P}},
    \end{gathered}\label{eq:RBF_PREVENT_OVERLAP}
\end{gather}
    where $D(g,i,n)$ determines if the RB $i$ of group $g$ contains RB fraction $n$ of the $N_{rf}$ total RB fractions. 
    
    The total transmission power of RB fractions in both layers is limited to  
    $P_{M}$, the maximum transmit power. The corresponding power allocation constraint is per \eqref{eq:RBF_MAX_PWR}.
\begin{gather}
    \begin{aligned}
     \sum_{c}^{C_{l,g}}{\sum_{i=0}^{L_g}{P_{min}(c,g)D(g,i,n)x_{g,l,c,i}}}  \le P_M\\
     l \in [0,1],  n \in [1,N_{rf}], \forall g \in \textbf{G}_p, p \in \bm{\mathcal{P}}
    \end{aligned}\label{eq:RBF_MAX_PWR}
\end{gather}
    where $P_{min}(c,g)$ gives the minimum transmission power required for group $g$ to achieve the lower bound SINR required for MCS $c$ per the SINR of the BS's pilot signal.  

\subsection{NOMA Inner Layer Optimization}
    NOMA inner layer optimization $I_p$ allocates resources specifically on the first layer of NOMA, i.e., when $l = 0$, per \eqref{eq:inner_NOMA}.
\begin{gather}
    x_{g,l,c,i} = 
    \begin{cases}
        1, & \exists g \in \textbf{G}_{p} \text{ and } l = 0\\
        0, & \text{otherwise}
    \end{cases} \label{eq:inner_NOMA}
\end{gather}
    By allocating the inner layer of NOMA, RBs can utilize the maximum transmit power from the BS. In other words, the BS transmits the QoS flow data with the best possible channel gain and MCS.  
        
    An allocated resource representation is shown in \eqref{eq:RB_FIX_OMA_HIGH_PRI}.
\begin{gather}
    \begin{gathered}
    R_{g,l,c} = \sum_{i=0}^{L_g}{x_{g,l,c,i}} ,  \\ 
    l \in [0,1], \forall c \in C_{l,g}, g \in \textbf{G}_p, p \in \bm{\mathcal{P}} \label{eq:RB_FIX_OMA_HIGH_PRI} 
    \end{gathered}
\end{gather}
    where $R_{g,l,c}$ are the time and frequency locations of the allocated RB of MCS $c$ of group $g$ in layer $l$. 

    RBs allocated to higher QoS priority groups $g \in \textbf{G}_{p'}$ where $p' < p$ (a lower $p$ has a higher priority) must be preserved while allocating resources to the lower QoS priority groups. 
    However, the assigned RBs' location may be re-assigned to different valid locations to utilize the radio resources fully.
    
    The ILP determines the inner layer's RB allocation. The inner layer's RB allocation forms the constraints for the following NOMA outer layer RB allocation $O_{p}$ solution space.
    
\subsection{NOMA Outer Layer Optimization}
    The NOMA outer layer optimization $O_p$ builds upon the inner layer's RB allocations. As such, it is similar to NOMA's inner layer allocation, yet its allocated RBs are specifically on NOMA's second layer. The assignment variable is defined per \eqref{eq:RB_FIX_NOMA_LOW_PRIO}.
    \begin{gather}
        x_{g,l,c,i} = 
        \begin{cases}
            1, & \exists g \in  \textbf{G}_p \text{ and } l = 1\\
            0, & \text{otherwise}
        \end{cases} \label{eq:RB_FIX_NOMA_LOW_PRIO}
    \end{gather}

    The NOMA outer layer allocation extends the allocated resources for high-priority groups (per \eqref{eq:RB_FIX_OMA_HIGH_PRI}) and the NOMA inner layer assigned RBs (per \eqref{eq:RB_FIX_OMA_HIGH_PRI} where $\forall g \in \textbf{G}_{p}$ and $l = 0$).
    
    Similar to inner NOMA resource allocation, location reassignment of RBs is allowed and enables combining similar geometries and spare power RBs to increase radio resource utilization.
    
    For the outer layer of NOMA RB allocation, multiple MCSs are considered due to the stochastic spare power coefficients resulting from the inner layer RB allocations.  
    The optimal solution derived through ILP is updated as additional constraints for the next NOMA inner layer resource allocation of the next QoS priority level $p+1$. 

\section{Approximation Algorithm}
\label{sec:Approx_Alg}
The optimal solution is NP-hard, as it is a combinatorial optimization problem. We can reduce the complexity by deriving an approximation algorithm that uses heuristics to reduce the search size of the solution space. 
For this research, we examine two numerologies. They are $0.5ms \cdot 360 kHz$ for high-priority QoS flows (5G RSU only supports high-priority QoS flows) and $0.5ms \cdot 180kHz$ for normal-priority QoS flows (4G BS supports high-priority and normal-priority QoS flows). We observe that mixing numerologies within an RB time-frequency location, which NOMA permits, should be prevented due to uneven interference \cite{INTER-NUM_INTF_5G_AND_BEYOND} generated by the low-latency resources onto the normal-latency resources. Incompatible RB geometries cause inefficient spectrum use. As such, we only permit similar RB configurations between guaranteed and non-guaranteed resources. Therefore, we pack low-latency QoS flows into as few RBs as possible to minimize the impact on normal-latency QoS flows. For 5G and NOMA, resource utilization is limited by the interference generated by non-similar RB geometries.

We propose using a maximum-benefit, minimum-harm heuristic to maximize RB utility. With this heuristic, in order of priority, RBs are allocated to each group based on which RB location provides the maximum gain for the group while minimizing the impact (interference) caused to other groups. Each device allocated RBs minimizes its transmit power while maintaining the same QoS afforded by its maximal power budget by using power control. RB power minimization minimizes the interference to other groups. Per this heuristic and the above observations, the candidate RB locations are limited to the unallocated RB locations. The number of candidate locations is upper-bounded by $N_{fr} \cdot L_g$, i.e., the number of possible $180 kHz$ aligned locations in each NOMA layer.  

We can allocate resources via the approximation algorithm shown in algorithm \ref{alg:NOMA_App}. The approximation algorithm is run once per radio sub-frame.  It assigns resources based on the best fit, per \eqref{eq:res_grad}. 
\begin{align}
	\label{eq:res_grad}
	&x_{g,l,c,i} = \begin{cases}
		\argmin\limits_x \left(c_{g,l,c,i} - D_g\right) \text{ if } \exists c_{g,l,c,i} \ge D_g \\
		\argmin\limits_x \left(D_g - c_{g,l,c,i}\right) \text{ otherwise} 
	\end{cases}
\end{align}
where $c_{g,l,c,i}$ represents the RB capacity for a specific RB location and $D_g$ corresponds to group $g$'s demand (bits).

\begin{algorithm}[t]
	\caption{Hybrid 4G/5G V2X (H45V) resource allocation approximation algorithm}
	\label{alg:NOMA_App}
	\begin{algorithmic}[1]
		\Require 
		\State $\textbf{P}$: list of QoS-prior social group flow sets.
		\Ensure \textbf{x} : RB assignments for each social group.
            \State $U$ = set of dual layer allocable RBs.\label{alg:QoS_Alloc_New}
		\State $S$ = set of outer layer allocable general RBs. 
		\State $H$ = set of outer layer allocable combinatorial RBs.  
		\State $D$ = set of dual layer allocated RBs.
		\For{$\textbf{G} \gets \textbf{P}$}
    		\State sort $G$ be ascending order of $g_\text{bits}/(g_\text{fair}*g_\text{size})$, where $g$ represents a social group setting. \label{alg:QoS_G_Demand}
    		\For{$g \gets \textbf{G}$} \label{alg:QoS_G_Start}
                    \State $S_\text{shape}$ := RBs in $S$ with shape $g_\text{shape}$. \label{alg:QoS_Alloc_Sym}
                    \State $\hat{S}_\text{shape}$ := RBs in $S$ except shape  $g_\text{shape}$. \label{alg:QoS_Alloc_Asym}      
                    \State $H_\text{shape}$ := RBs in $H$ with outer layer shape $g_\text{shape}$. \label{alg:QoS_Alloc_Comb} 
                    \For {$R \gets [S_\text{shape},H_\text{shape},U]$}
    		        \State sort $R$ to ascend the remaining transmit power.
    		        \For{$r \gets R$}
                        \If{$g_\text{bits} > 0$}
    		                \State remove $r$ from $R$.
                                \State decrease $g_\text{bits}$ with allocated bits.
    		                \State add $r':= r + g$ into $D$.
                        \EndIf
    		        \EndFor
                    \EndFor
                    
    		    \If {RB to satisfy $g_\text{shape} \le |\hat{S}_\text{shape}|$}
    		        \State sort $\hat{S}_\text{shape}$ to ascend the remaining transmit power.
    		        \For{$r \gets S_\text{shape}'$}
                        \If{$g_\text{bits} > 0$}
    		                \State remove $r$ from S
                                \State decrease $g_\text{bits}$ with allocated bits.
                                \State add $r':= r + g$ into $H$.
                        \EndIf
    		        \EndFor
    		    \EndIf
    		\EndFor 
    		 \label{alg:QoS_NG_Demand}
		\EndFor \label{alg:QoS_G_End}
	\end{algorithmic}
\end{algorithm} 

For each set of prioritized QoS group flows $\mathbf{G}$, we sort the flows by the group's demand, fairness, and size on line \ref{alg:QoS_G_Demand}. Then, on lines \ref{alg:QoS_G_Start}-\ref{alg:QoS_G_End}, QoS flows are allocated resources. We first consider RB locations that have allocated the inner layer resources to groups of identical RB shape $g_\text{shape}$ (line \ref{alg:QoS_Alloc_Sym}). Assuming a suitable fit is possible, we assign the QoS flow to the tightest fit possible, as the assignment minimizes the impact on the remaining flows. 
After which, we consider spare outer layer combinatorial RBs (line \ref{alg:QoS_Alloc_Asym}). These RBs have inner layer RBs with different shapes than the outer layer RB. An example of spare combinatorial RBs with a 360 kHz$\cdot$0.5 ms RB allocable in the outer layer is one consisting of two 180 kHz$\cdot$1 ms RBs forming a 360 kHz$\cdot$1 ms RB in the inner layer and a 360 kHz$\cdot$0.5 ms RB in the outer layer. Then, we select non-allocated RBs (line \ref{alg:QoS_Alloc_New}) to allocate in an impact-minimizing manner. The non-allocated RBs are selected at locations with neighboring RBs having identical RB shapes. Finally, we select non-identical shape inner layer RBs containing spare outer layer resources to form a combinatorial RB (line \ref{alg:QoS_Alloc_Comb}), then we allocate part of the resources in the outer layer of the combined RB. The selection process is performed in an impact-minimizing manner, i.e., we minimize the number of RBs required to satisfy the outer layer's resource demands. 

\subsection{Computational complexity}
The optimal solution to the resource allocation problem is NP-hard, as all feasible resource allocations and power settings must be investigated. Our approximation algorithm, on the other hand, has a polynomial run time. We define computational complexity in terms of the number of QoS priorities, $\bf{\mathcal{P}}$, and the number of social groups $\mathbf{G}$ at a set QoS priority $p$.  
First, each set of QoS flows is sorted by its rate gradient $g_{bits} / (g_{fair} \cdot g_{size})$. Sorting requires at most $O(MlogM)$, where $M = \max\limits_{p \in \mathbf{G}} |\mathbf{G}_p|$. Then the resources are assigned in a max-min manner requiring $O(3 \mathbf{G} M log(\mathbf{G} M) + 1)$ time, as each flow must inspect three distinct sets of possible RB locations for other QoS flows, given the RB configuration. Therefore, the time required for our approximation algorithms is: $O(\mathbf{G} (M log M + M(3 \mathbf{G} M log(\mathbf{G} M) + 1))) \approx O((\mathbf{G} M)^2 log(\mathbf{G} M))$.

\section{Simulation and Discussion}
\label{sec:Sim_Dis}
We simulated vehicular traffic in the area (street map) surrounding National Taiwan University of Science and Technology (NTUST), Taiwan, by utilizing SUMO\cite{SUMO2018}. 
\begin{figure}[t]
	\centering
	\begin{minipage}{1.0\columnwidth}
	\subfloat[BS + RSU view]{
	 \frame{\includegraphics[trim=3mm 0mm 5mm 15mm, clip, width=0.49\columnwidth]{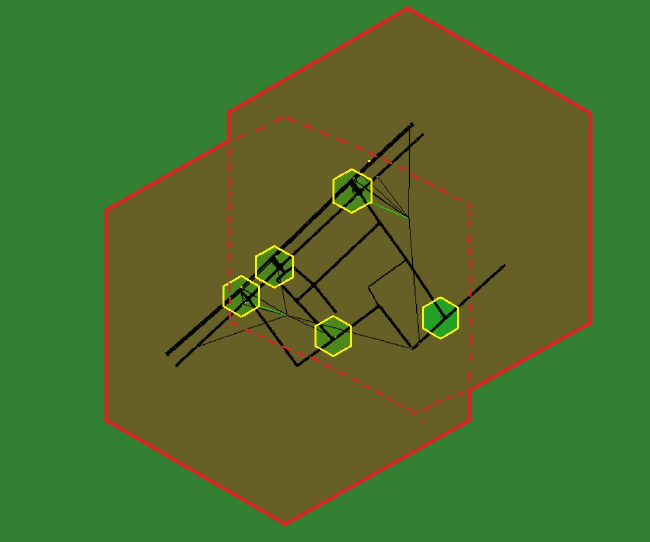}}
	 \label{fig:SUMO_Sim_Macro}
	}
	\subfloat[RSU view]{
		\frame{\includegraphics[trim=3mm 0mm 5mm 15mm, clip,width=0.47\columnwidth]{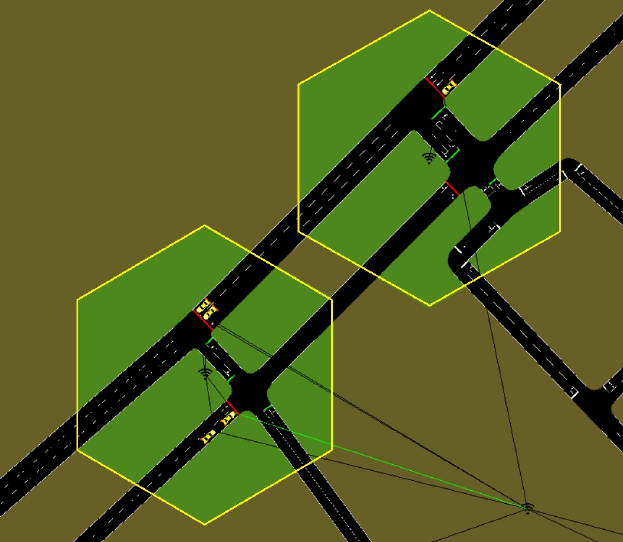}}
		\label{fig:SUMO_Sim_Small}
	}
	\end{minipage}
	
	\caption{National Taiwan University Science and Technology SUMO Simulation: macrocells are colored red, while RSUs are colored green.}
	\label{fig:SUMO_Sim}
\end{figure}
In our simulation, we assumed a Poisson packet arrival as determined by the network traffic type and the QoS flow/packet size. We utilized the 3GPP radio models per TR 38.901\cite{TR_38901}. RSUs appear as green cells, and BSs appear as red cells with locations as per Fig. \ref{fig:SUMO_Sim}. The bandwidth allocated to both the BSs and each RSU is split symmetrically per BS/RSU between uplink and downlink. The simulation parameters we used are shown in table \ref{tbl:Sim_Params}.  The CQI to SINR mapping is per \cite{TS_36213}. 
\begin{table}[b]
	\caption{Simulation Parameters}
	\centering
	\label{tbl:Sim_Params}
	\begin{tabular}{|l|l|l|} 
		\hline \textbf{Parameter} & \textbf{Value} \\
		\hline Number of Simulations & 20 \\
		\hline Simulation Duration & 10 seconds \\
		\hline Average Number of Vehicles & $100 \to 140$ \\ 
		\hline CQI update rate & 100 ms \\
		\hline BS/RSU Bandwidth (symmetric) & 20 Mhz /10 Mhz \\
		\hline BS/RSU Tx power & 23 dBm / 10 dBm \\
		\hline BS/RSU Radio model & UMa / UMi \\
		\hline BS/RSU Frequency & 2.0 GHz / 3.5 GHz\\
		\hline BS/RSU Range: $D_M$ / $D_R$ & 500m / 50m \\
		\hline Number of BSs/RSUs: $|B|$ /  $|R|$ & 2 / 5 \\
		\hline Critical, High-priority, QoS flow size $(\mathcal{F})$ & $\mathcal{U}$(300-1100) bytes \\
		\hline Critical, High-Priority, average flow size $(\overline{\mathcal{F}})$ & $700$ bytes\\
		\hline Critical, High-Priority, average request rate $(\mathcal{R})$ & $11$ pps\tablefootnote{packets per second}\\
		\hline Critical, High-Priority, average bit rate $(\mathcal{D})$ & $64$ kbps \\
		\hline General, Normal-Priority, QoS flow size $(\mathcal{F})$ & $\mathcal{U}$(64-2048) bytes \\
		\hline General, Normal-Priority, average flow size $(\overline{\mathcal{F}})$ & $1056$ bytes \\
		\hline General, Normal-Priority, average request rate $(\mathcal{R})$ & $60$ pps \\
		\hline General, Normal-Priority, average bit rate $(\mathcal{D})$ & $500$ kbps \\
        \hline Fairness time constant $T_c$ & 300 ms \\
		\hline
	\end{tabular}
\end{table}

For our simulations, the relationship between the number of groups ($N_G$), max group size ($S_{G, max}$), and number of vehicles ($N_V$) are per \eqref{eq:Sim_Sizes}.  
\begin{equation}
    \label{eq:Sim_Sizes}
    N_G \ge \frac{N_V}{S_{G,max}}
\end{equation} 
Unless otherwise stated, we fix the maximum group size $S_{G, max}$ at 20. The size of each group dynamically changes throughout each simulation as vehicles move in/out of the range of each BS and RSU. In all simulations, as only critical high-priority RBs may use RSUs, each 4G macrocell BS only supports 1ms x 180 kHz RB geometry, while each 5G RSU only supports 0.5ms x 360 kHz RB geometry.

For our simulations, we evaluate the following metrics:
\begin{itemize} 
    \item Average Transfer Time: The duration of time a QoS flow occupies radio resources.
    \item Average End-to-End Time: The time from a QoS flow's first packet generation until its last packet reception, including preemption time.    
    \item Average Timeout Ratio: The ratio of QoS flows starved of resources per \eqref{eq:TO_Ratio}, i.e., the receiver has not received the last packet within seven seconds of the initial uplink queuing time of the QoS flow. 
    \item Average Queue Time: The time a QoS flow is en-queued in the system, i.e., the transmission has started, yet it is waiting for resources to complete.
\end{itemize}

\noindent The timeout ratio $T_q$ is specified per \eqref{eq:TO_Ratio}:
\begin{align}
        \label{eq:TO_Ratio}
        T_{q} = \frac{|F_q^T|}{|F_{Q}|},\ 
    	F_{q}^{T} = 
    	\left\{
             d \in F_{Q} \ 
            \begin{tabular}{|l}
              $d_t > 7s$
            \end{tabular}
        \right\}   
\end{align}
\noindent where the QoS setting is $q$, $F_q^T$ represents the set of QoS flows transmitted. $F_{Q}$ represents all QoS flows, and $d_t$ is the duration of a network flow from the initial transmission time (first packet) to the end of its reception (last packet).  

As our resource allocation methodology is QoS-aware, high-priority flows are allocated resources prior to normal-priority flows, i.e., throughput is not our primary consideration among QoS classes as throughput is inherently unfair to QoS flows with lower demand, potentially causing starvation.  In this work, we derive a fair resource allocation approach that ensures each QoS flow is eventually allowed to transmit its data, thus reducing the likelihood of starvation.  As a result, our mechanism lowers the overall latency by allowing each QoS flow to transmit.

In all simulations, we compare general (normal-priority QoS flows) and critical (high-priority QoS flows) network traffic flows.   The simulated radio conditions demonstrate how channel gains, corresponding queuing duration, and network traffic flows impact the system's QoS when considering vehicle location. 

Compared to BS-only and socially aware deployments, the strength of QoS reclassify V2X is that when critical/high-priority traffic is not urgent to non-local BS users, as is the case where the traffic originates from an RSU and forwarded to a different locality. When the traffic is forwarded, the BS reclassifies the traffic as normal-priority traffic, i.e., the BS minimizes the QoS impact of high-priority traffic on normal-priority traffic. However, when RCWS traffic is considered urgent to other vehicles (vehicles within the range of the RSU/BS receiving the high-priority V2X network flow), the QoS flow's priority remains unaltered within the originating locality. Using the original QoS priority ensures timely delivery to the vehicles near the scene of incidence. 
If members of the same target group associate with different BSs, high-priority network traffic within a BS is forwarded to other macrocells, without QoS reclassification, to support group member communications.

In our bi-level NOMA system, high-priority connections only use the inner-layer NOMA, while normal-priority connections use either layer of NOMA. 
Network traffic with a target group associated with a BS  / RSU is groupcast to the group members within range of the BS  / RSU. Safety-related RCWS social groups include all vehicles as members, i.e., RCWS traffic broadcasts to all vehicles. 
Non-safety groups consider group member conditions, i.e., network traffic is groupcast. The following sections discuss and analyze broadcast and groupcast traffic, the system's queuing state, and the QoS Impact.

\begin{figure*}[t]
    \centering
    \begin{minipage}{0.495\textwidth}    
        \subfloat[Generic QoS Flows]{\includegraphics[width=0.495\columnwidth]{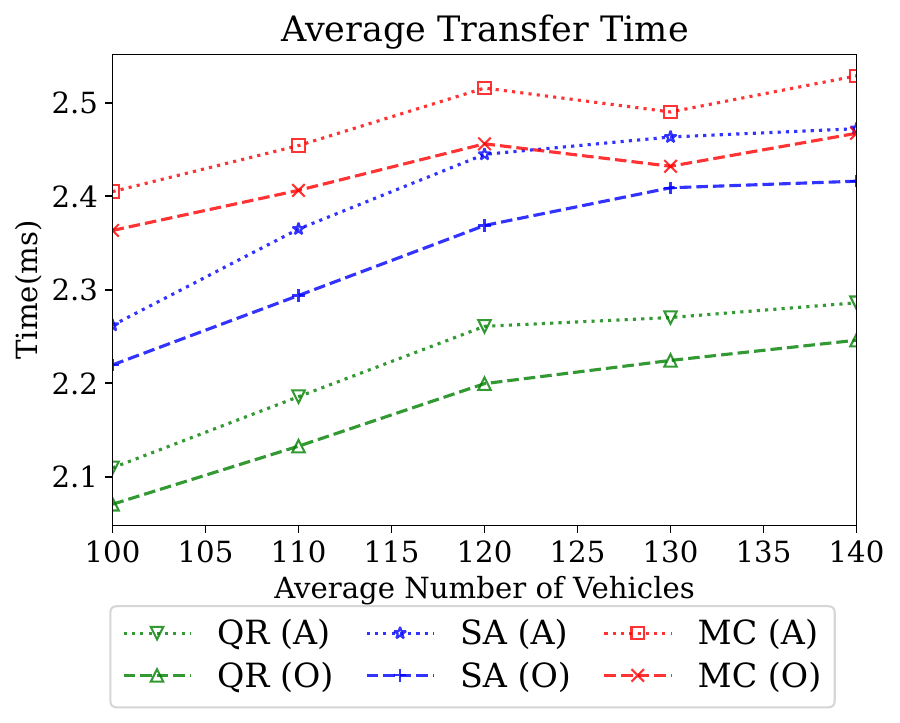}}\label{fig:TT_GNR} 
        \subfloat[Critical QoS Flows]{\includegraphics[width=0.495\columnwidth]{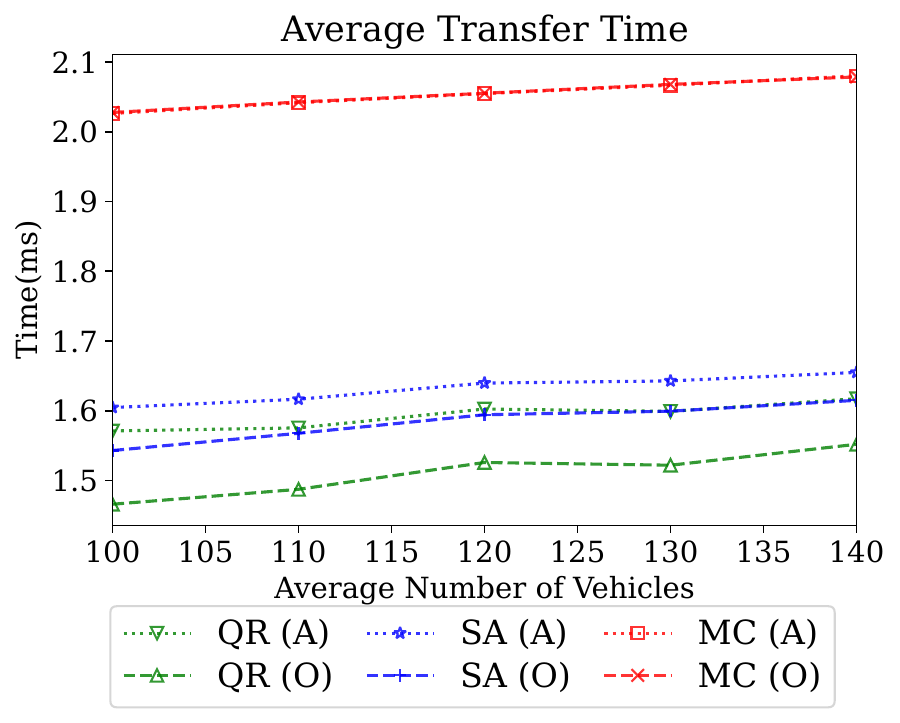}}\label{fig:TT_CR} 
        \caption{Average Transfer Time}
        \label{fig:TT} 
    \end{minipage}
    \begin{minipage}{0.495\textwidth}
        \subfloat[Generic QoS Flows]{\includegraphics[width=0.495\columnwidth]{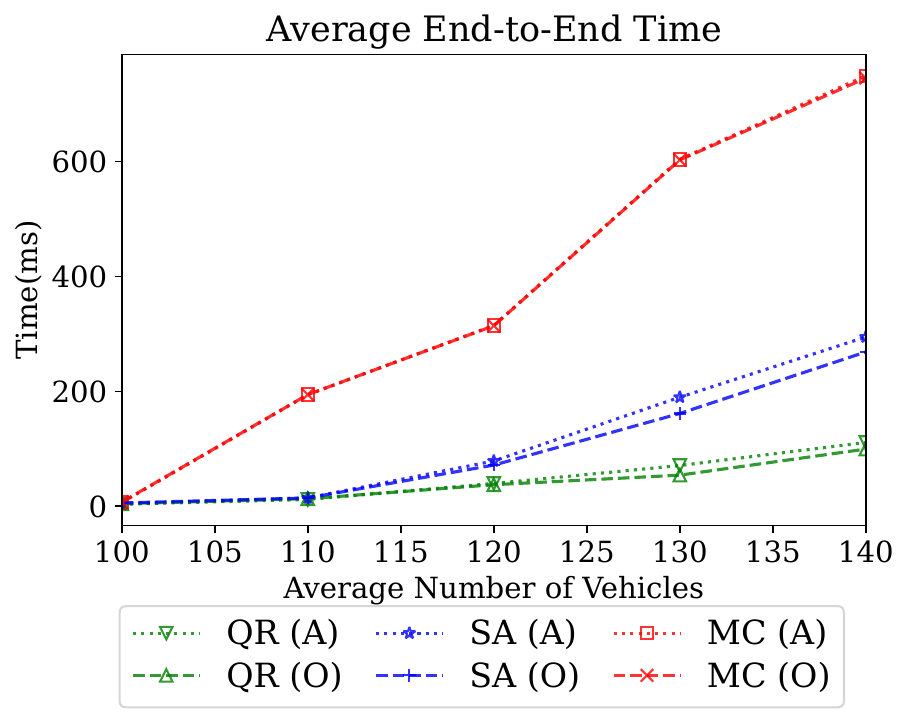}}\label{fig:E2E_GNR} 
        \subfloat[Critical QoS Flows]{\includegraphics[width=0.495\columnwidth]{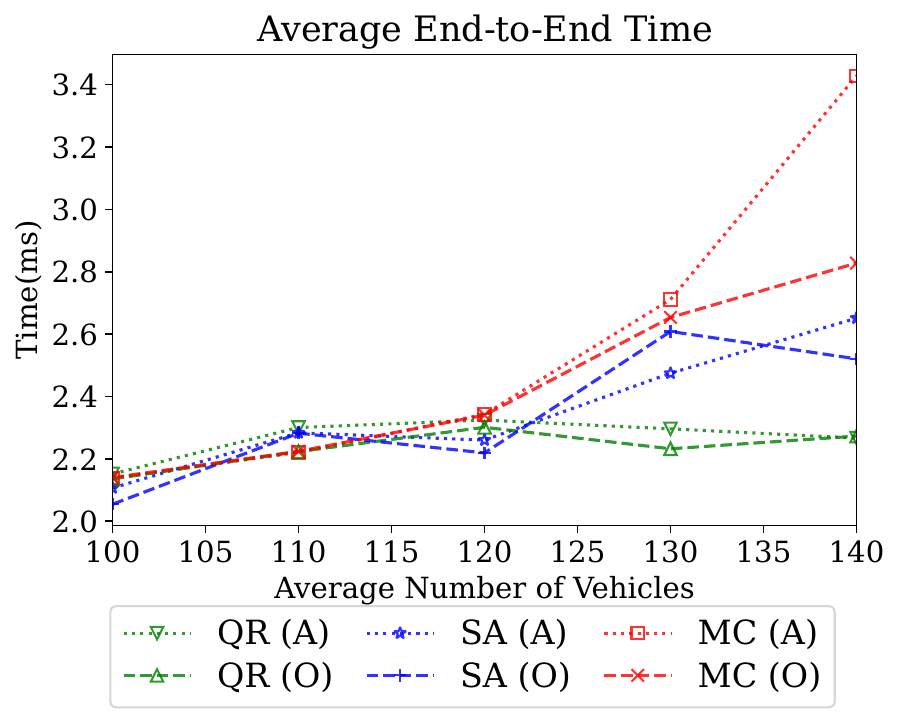}}\label{fig:E2E_CR} 
        \caption{Average End-to-End Time}
        \label{fig:E2E}     
    \end{minipage}
\end{figure*}    
\begin{figure*}[t]
    \centering
    \begin{minipage}{0.495\textwidth} 
        \subfloat[Generic QoS Flows]{\includegraphics[width=0.495\columnwidth]{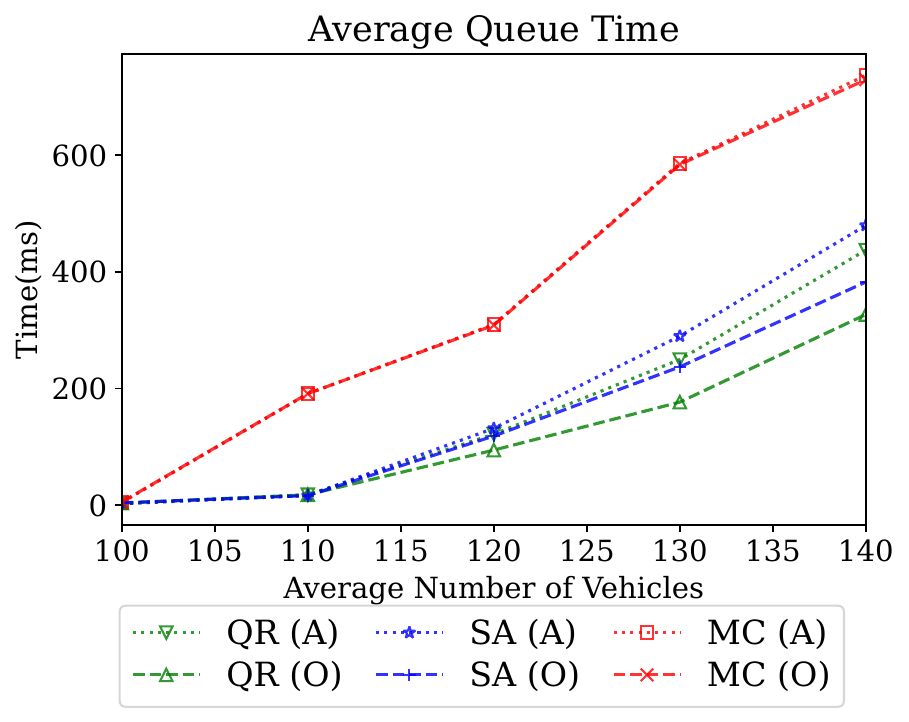}}\label{fig:QT_GNR} 
        \subfloat[Critical QoS Flows]{\includegraphics[width=0.495\columnwidth]{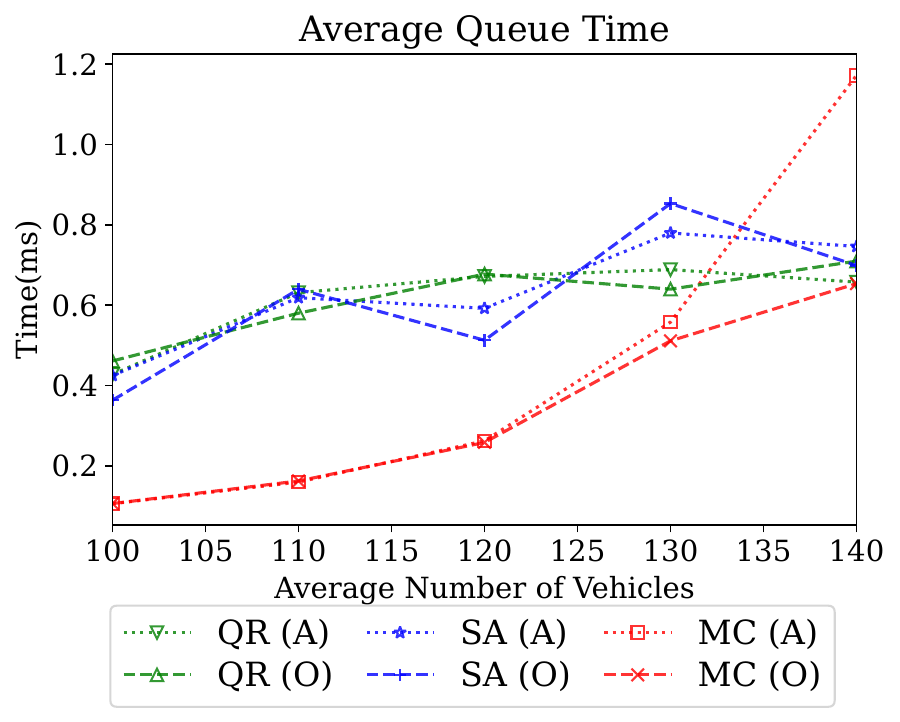}}\label{fig:QT_CR} 
        \caption{Average Queue Time}
        \label{fig:QT} 
    \end{minipage}
    \begin{minipage}{0.495\textwidth} 
        \subfloat[Generic QoS Flows]{\includegraphics[width=0.495\columnwidth]{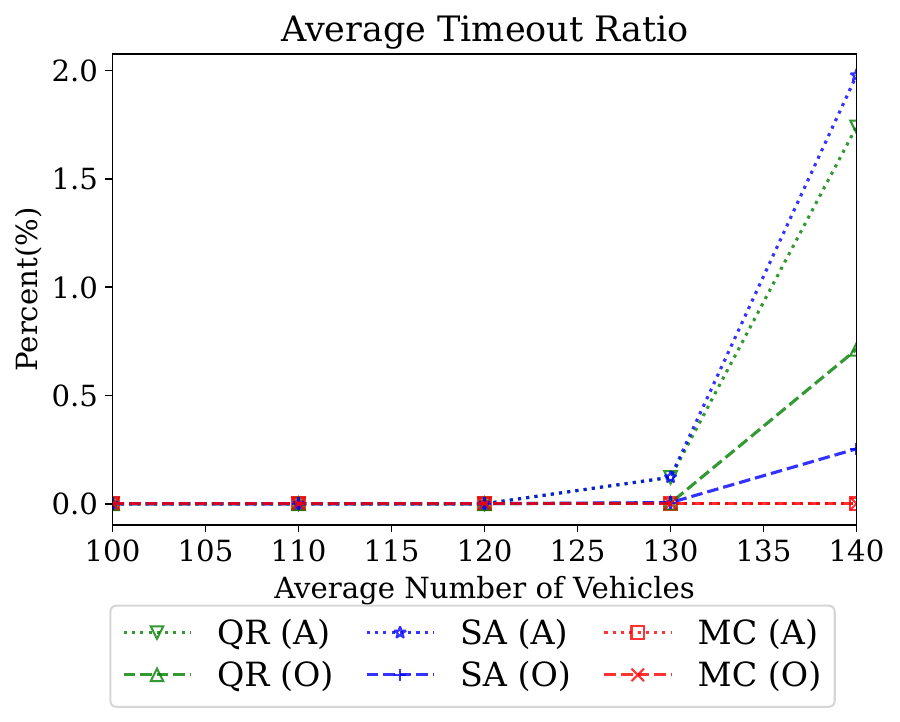}}\label{fig:TR_GNR}
        \subfloat[Critical QoS Flows]{\includegraphics[width=0.495\columnwidth]{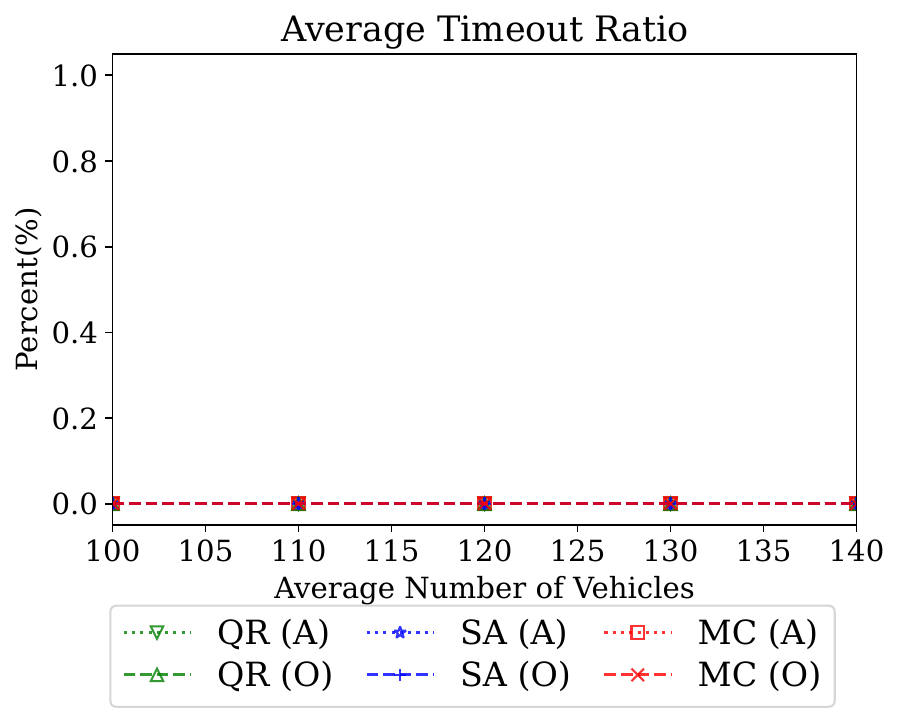}}\label{fig:TR_CR}
        \caption{Average Timeout Ratio}
        \label{fig:TR}  
    \end{minipage}
\end{figure*}

In figures \ref{fig:TT} $\to$ \ref{fig:TR},
we show simulations for three different deployments: NOMA-enabled 4G macrocell BS only V2X ("MC" series data), socially aware 5G V2X ("SA" series data), and QoS reclassifying 5G V2X ("QR" series data). For each deployment, we analyze the general and critical QoS flows under both the optimal resource allocation algorithm (data series includes "(O)") and the approximation resource allocation algorithm (data series includes "(A)"). 

In the QoS reclassify deployment, data that started and ended as critical high-priority data impacts critical QoS flows. Data that started as critical and the BS reclassified general impacts both critical QoS flows and general QoS flows. Finally, data that started as normal QoS only impacts general QoS flows. 

\subsection{Analysis of fixed group parameters}
In this section, we consider a fixed maximum group size to understand better how the number of vehicles impacts the system's behavior.
We first analyze the average transfer time.  
Across all deployments in Fig. \ref{fig:TT}, as the volume of high-priority (critical) traffic increases, the normal-priority (general) traffic is preempted more often. When no RSU is present / in range, the high-priority traffic preempts uplink and downlinks region-wide BS resources. In all cases, the approximation algorithm requires at least as much time as the optimal algorithm. The optimal algorithm simulation acts as a lower bound on the transfer time.
In the socially aware deployment, the benefit of socially aware V2X is evident via the reduced transfer time due to RSU resource utilization by critical QoS uplink/downlink. The RSUs provide additional uplink radio resources explicitly provided for critical network flows. 
Note:  When an RSU forwards data to a BS, the critical QoS flow transfer times are aggregated and averaged. The aggregate transfer time = RSU uplink + RSU downlink (local transmission) + RSU uplink + BS downlink (regional transmission).
In the QoS reclassify deployment, we observe a decrease in transfer time relative to socially aware as fewer critical flows compete for resources. Thus, more resources are assignable to general network flows.
For general QoS flows, QoS reclassify V2X shows a lower average transfer time as critical flows forwarded to the BS are reclassified as general flows, i.e., no unnecessary preemption occurs. As critical flows remain critical within the hierarchy they originate from, critical transfer time decreases due to said flows being QoS reclassified when forwarded.   As QoS flows share the spectrum, fewer critical network flows compete for the same radio resources.
In general, when comparing the 4G BS (only) to the 4G BS + 5G RSU configurations, the differences in transfer time are primarily due to the RB geometry and the additional RSU resources. While the approximation algorithm does not achieve as tight results as the optimal algorithm, the differences are negligible.  
Comparing the generic QoS flow series across all three deployments, our proposed QoS reclassify V2X scheme's end-to-end time requires $\approx 15\%$ less time than 4G BS (only) and $\approx 10\%$ less time required by a socially aware 5G V2X deployment.

The average QoS flow end-to-end time is displayed in Fig. \ref{fig:E2E}. Per the 4G BS (only) deployment, as only a BS is present, critical traffic is serviced first, while general traffic incurs delay due to preemption. When we consider QoS reclassifying V2X, we notice both socially aware and QoS reclassify 4G BS + 5G RSU schemes outperform 4G BS (only). However, we also notice that socially aware V2X performs worse than QoS reclassifying V2X when traffic congestion occurs, starting from 120 vehicles. In socially aware V2X, high-priority traffic is forwarded from the RSU to the BS. The QoS is unchanged, as data relevance is not a factor. Consequently, RSU originating high-priority traffic preempts normal-priority traffic on the BS. Compared to QoS reclassify V2X, there is a higher end-to-end time for the general QoS flows, as is evident from 120 vehicles.
In the QoS reclassify V2X, the normal-priority QoS general traffic 
does not incur any unnecessary delay, i.e., the critical traffic does not preempt the normal traffic as it is reclassified as normal-priority QoS. 
We note that the approximation algorithm performs similarly to the theoretically optimal results. Comparing the generic QoS flow series across all three deployments, in terms of transmitting a QoS flow, our proposed QoS reclassify V2X scheme's end-to-end time requires only $\approx 15\%$ of the time required by a 4G BS (only) and $\approx 30\%$ of time required of a socially aware 5G V2X deployment.

We now consider the queuing time graphs, as shown in Fig. \ref{fig:QT}.  
In the 4G BS (only) deployment, we observe that the queuing time is nearly linear and almost identical for critical and general traffic due to the preemption of critical traffic. 
In the socially aware deployment, we observe a lower queuing time than 4G macrocell BS only as local to the RSU flows do not need to compete for regional resources.
Finally, in our QoS reclassify deployment, we observe that the queuing time slope is lower than the socially aware deployments. The lower queueing time is due to general latency QoS flows not being preempted by inbound inter-BS/RSU traffic. 

We now look at the timeout ratio. Fig. \ref{fig:TR} shows the QoS flow timeout ratio graphs.  
Timeout can occur on a QoS flow if it generates network traffic, yet the receiver only partially receives the QoS flow within seven seconds from the QoS flow’s generation. In other words, a timeout occurs if the QoS flow queuing time plus transmission time exceeds seven seconds. In our setup, critical high-priority traffic is local to either the RSU or BS of origin. If a critical QoS flow incurred a timeout, only the critical QoS flow would reflect this. General traffic was either critical traffic that was duplicated and reclassified or started as general traffic. If a general QoS flow incurred a timeout, only the general QoS flow would reflect this.  

First, looking into the BS-only deployment, we see that the timeout ratio is relatively low, less than $\frac{1}{100}^{th}\%$. The implication is that the bottleneck is in radio access, as the amount of resources allocated for uplink is identical to that of the downlink. 
A BS-only deployment alleviates the timeout ratio, as transmissions need to queue for resource allocation, but sacrifices the gain obtained by priority groups during downlink resource allocation as the BS uplink and broadcast receive all RCWS-generated traffic to the cell. Looking into the socially aware V2X deployment, the BSs receive QoS flows forwarded by the RSUs.  As downlink allocation/demand is no longer identical to uplink, the resource allocation algorithm allocates downlink resources to higher-priority groups. As such, socially aware suffers from higher demand, longer queuing times, and a noticeable timeout ratio.
Finally, we examine the QoS reclassified V2X deployment where high-priority flows are reclassified as a normal-priority before being forwarded. We note that QoS reclassification has a lower timeout ratio than socially aware. Also, we observe that the approximation algorithm performs similarly to the theoretically optimal algorithm until network congestion occurs around $120$ vehicles, after which the deviation between both methods becomes more apparent.
Relative to socially aware, the lower timeout ratio achieved by the 4G macrocell BS only deployment is due to an increased number of serviced high-priority QoS flows.  
Also, we observe that the QoS reclassify timeout ratio has increased relative to the socially aware due to high-priority flow reclassification. The increase in normal-priority traffic generates a higher timeout ratio. The optimal and approximation timeout ratios converge if critical flows increase. 
Until network congestion occurs, our approximation algorithm performs similarly to the optimal algorithm.

\subsection{QoS Impact Analysis}
Ultimately, the queuing state of the system determines the impact on each QoS flow. When a vehicle or multiple BSs (inter-BS) are queuing for resources due to preemption or the data arrival rate at the BS exceeding its service rate, the queuing time increases. The cause ultimately determines the extent of the queue increase.  
We note that the consequences of increased queuing time are longer end-to-end times and a higher probability of a timeout event, i.e., a higher timeout ratio.  

When studying the impact on normal-priority network flows, the shift from sufficient to insufficient wireless resources becomes evident. Per Fig. \ref{fig:QT}, the 4G BS (only) and socially aware deployments are noticeably impacted. When wireless resources are insufficient, the average queuing time is higher. Higher queuing time is due to network traffic queuing for resources, i.e., higher competition.  

When considering socially aware V2X networks and comparing them to 4G V2X, we find that the addition of 5G RSUs results in a reduction in the average queuing time. The network traffic maintains the same QoS priority when the RSU/BS forwards data to other BSs yet has additional uplink resources to exploit. 
In socially aware networks, wireless resource competition decreases due to vehicles directing high-priority traffic to RSUs (when available), and inter-BS traffic increases due to network flow duplication. Note: BS originating and RSU-destined traffic maintains its original QoS. 
Thus, the average queuing time becomes dominated by the inter-BS queuing delay.  

Finally, when considering QoS reclassifying V2X compared to socially aware V2X networks, we see a further reduction in queuing time as not only are high-priority network flows being directed to RSUs, but they are also being reclassified as normal "non-preempting" priority QoS flows. Thus, lower wireless resource competition and a lower inter-BS preemption rate result in an even lower average queuing time.

\section{Conclusion}
\label{Sec:Conclusion}
This research investigated how social networks can be combined with MEC servers and DC to deploy localized QoS for socially aware V2X networks. In our proposed system, the primary connection is a 4G BS, and the secondary connection is a 5G RSU. Our system ensures compatibility with existing V2X deployments. We found that each vehicle's social network (origin and target BS/RSU) and traffic type can define the QoS setting for each vehicle's associated QoS flows. The benefit of our socially aware deployment is two-fold: 1) localized QoS is supported by leveraging the underlying social network between vehicles and their QoS flows, and 2) legacy devices (4G) can inherently use our system without being aware of its presence.

By reclassifying critical traffic as general traffic, our socially aware DC system reduces our V2X system's overall QoS flow end-to-end time, queuing duration, and timeout ratio without adversely impacting critical traffic's QoS per target audience. This results from each QoS flow's localized QoS priority differing from its regional QoS priority, i.e., the QoS priority for critical network traffic remains unaltered within the vicinity of the event yet is modified when forwarded for regional distribution.

\section{Future Work}
\label{sec:Future_Work}
In traditional cellular networks, a network traffic's QoS priority selection is configured when a data flow is created. However, static QoS definitions ultimately lead to inefficient resource utilization. Nevertheless, we can improve resource allocation and utilization in near real-time via Artificial Intelligence (AI) on the network (6G/5G and AI). 
In our future work, we plan to extend QoS reclassification to network-wide AI-managed QoS reclassification based on social groups and each network interface's traffic type, origin, destination, and capacity.

\bibliographystyle{IEEEtran}
\bibliography{IEEEabrv, Refs/V2X_Ref, Refs/PF_NOMA_Schedule, Refs/QoS-aware}
\end{document}